%% file: iclr2025_conference.tex
\documentclass{article} 
\usepackage{iclr2025_conference,times}

\input{math_commands.tex}

\usepackage{hyperref}
\usepackage{url}
\usepackage{graphicx}
\usepackage{enumitem}
\usepackage{booktabs}
\usepackage{subfigure}
\usepackage{multirow}
\usepackage{caption}

\title{
Towards Generalized Routing: Model and Agent Orchestration for Adaptive and Efficient Inference}

\author{
Xiyu Guo, Shan Wang, Chunfang Ji, Xuefeng Zhao$^*$, 
Wenhao Xi, Yaoyao Liu, Qinglan Li,  \\
\textbf{ Chao Deng, Junlan Feng\thanks{ The corresponding author.}}
\\
JIUTIAN Team, China Mobile Research Institute, Beijing, China\\
\texttt{\{zhaoxuefeng,fengjunlan\}@chinamobile.com} \\
}

\begin{document}

\maketitle

\begin{abstract}
The rapid advancement of large language models (LLMs) and domain-specific AI agents has greatly expanded the ecosystem of AI-powered services. 
User queries, however, are highly diverse and often span multiple domains and task types, resulting in a complex and heterogeneous landscape. 
This diversity presents a fundamental routing challenge: how to accurately direct each query to an appropriate execution unit while optimizing both performance and efficiency.
To address this, we propose MoMA (Mixture of Models and Agents), a generalized routing framework that integrates both LLM and agent-based routing. 
Built upon a deep understanding of model and agent capabilities, MoMA effectively handles diverse queries through precise intent recognition and adaptive routing strategies, achieving an optimal balance between efficiency and cost.
Specifically, we construct a detailed training dataset to profile the capabilities of various LLMs under different routing model structures, identifying the most suitable tasks for each LLM. During inference, queries are dynamically routed to the LLM with the best cost-performance efficiency. 
We also introduce an efficient agent selection strategy based on a context-aware state machine and dynamic masking.
Experimental results demonstrate that the MoMA router offers superior cost-efficiency and scalability compared to existing approaches.

\end{abstract}

\section{Introduction}
In recent years, the ecosystem of LLMs and AI agents has grown at an unprecedented pace, giving rise to a diverse spectrum of systems with different resource demands, domain expertise, and reasoning paradigms.
Representative examples include general-purpose LLMs such as GPT-5 \footnote{https://openai.com/index/introducing-gpt-5/ \label{ft1}}, domain-specific models like Med-PaLM \citep{medpaLLM} for medical applications, as well as specialized agents such as Cursor Agent for code generation \citep{curson} or JoyAgent for e-commerce services \citep{joyagents}.
At the same time, user queries themselves are highly heterogeneous. 
A capability-aware matching strategy is typically employed. 
Specialized and complex tasks, involving tool invocation, multi-step reasoning, or long-horizon planning, are better suited for agent-based solutions. More straightforward tasks like knowledge retrieval or text generation are handled by general-purpose LLMs. 
As a result, relying exclusively on either LLMs or agents is inadequate for covering the full spectrum of real-world scenarios. This leads to a fundamental challenge: \textbf{how can we efficiently and reliably select the most appropriate execution unit from a heterogeneous pool of models and agents to deliver robust and cost-effective adaptive services?}


\begin{figure}[ht]
  \centering
  \includegraphics[width=13.8cm]{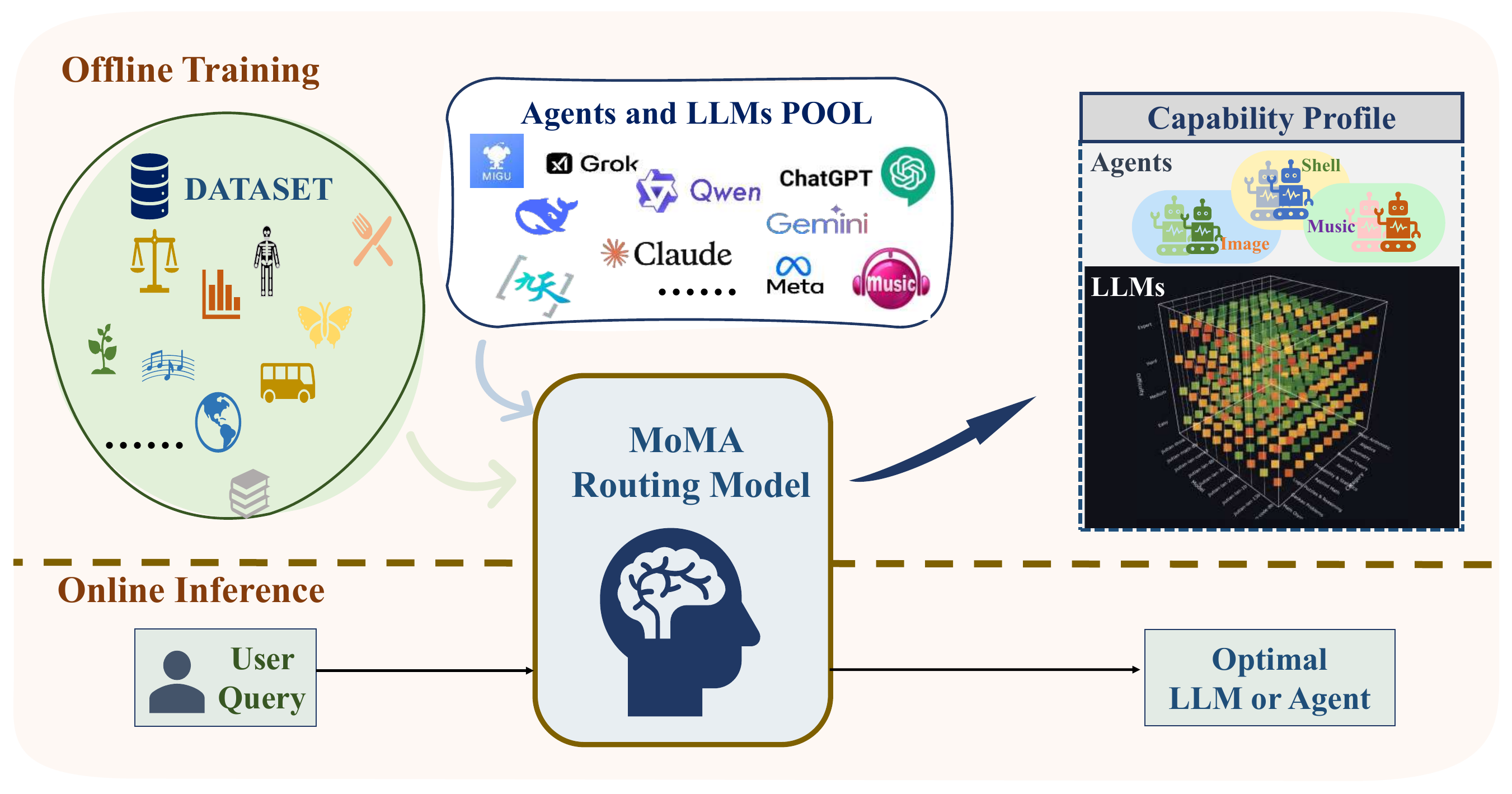}
  \caption{Illustration of the proposed adaptive routing model.}
  \label{intro}
\end{figure}

This paper aims to develop an adaptive and generalized routing model, as shown in Figure~\ref{intro}.
During the training phase, the routing model learns from the constructed large-scale and extensive dataset, incorporating LLMs and agents within the resource pool, ultimately effectively characterizing the capabilities of both LLMs and agents across various domains. 
During the inference phase, the trained routing model leverages learned knowledge to accurately map user queries to the most appropriate agent or LLM for response.

Some works focusing only on LLMs routing highlight a fundamental trade-off between performance and efficiency: lightweight LLMs offer lower computational costs and latency but suffer from limited reasoning and generation capabilities. 
Several approaches have been proposed to address this challenge. For example, GPT-5’s router \footref{ft1} dynamically assigns each query to an appropriate model to balance performance and efficiency.
RouterLLM \citep{ong2024routellm} trains a binary classifier using preference data to route queries to stronger or weaker models.
In addition, RouterDC \citep{chen2024routerdc} leverages dual contrastive learning to improve routing accuracy. 
While these methods achieve certain performance–cost trade-offs, they generally target only a small number of pre-specified models and struggle to scale to a heterogeneous LLM pool with diverse parameter sizes and continuously growing numbers, leading to limited adaptability.
AvengersPro \citep{zhang2025beyond} embeds and clusters queries, routing them to LLMs based on performance–efficiency scores. However, this approach lacks training for a dedicated routing model, relying on a coarse-grained matching to link user queries with LLMs, which cannot accurately assess the LLM's performance across different user queries. 
What's more, recent research on multi-agent systems has also revealed promising directions. The Mixture of Agents (MoA) \citep{wang2024mixture} architecture surpasses GPT-4 Omni by leveraging multi-round interactions among a set of medium-sized models (70B-level parameters). Building on this, variants such as sparse MoA \citep{sparseMOA} and Self MoA \citep{selfMOA} have been introduced. However, it remains a pivotal and critical issue to accurately and efficiently invoke agents based on task features.



\textbf{Our work is the first to present a generalized routing model that jointly considers LLM and agent routing to effectively handle a wide range of heterogeneous user queries}, which face several major challenges. 
First, it is far from trivial to characterize the LLM profile, especially when facing LLMs from similar domains, which places stringent demands on the construction and augmentation of the dataset.
Moreover, designing a routing model that achieves accurate orchestration and cost-efficient inference, while effectively harnessing the potential of an expanding and heterogeneous model pool, remains a formidable challenge.
Furthermore, the expansion of the agent ecosystem complicates precise intent-agent matching due to increasing functional overlaps.



To this end, we propose a routing model, Mixtures of Models and Agents (MoMA), to deliver large-scale and diverse services under cost–performance trade-offs. 
Drawing upon a profound understanding of model and agent capabilities, MoMA employs precise intent recognition and adaptive routing strategies to not only align user queries with the most suitable execution unit but also optimize routing efficiency and cost-effectiveness.
The main contributions of this paper are summarized as follows:
\begin{itemize}
\item \textbf{Framework}: 
We are the first to unify routing across multiple LLMs and agents, enabling real-time and dynamic scheduling based on user queries. This integration builds a more robust and adaptive solution for diverse and complex tasks.

\item \textbf{Router Design}: 
We train a router by meticulously constructing the training dataset and designing the model structure to adaptively match user queries to the most suitable execution unit, aiming to achieve a balance between inference performance and user cost for each request by leveraging Pareto-optimal principles.

\item \textbf{Exploring LLMs Capability}: 
We explore and analyse the performance of LLMs across a range of parameter scales tailored to specific task requirements, revealing the inference potential of various models, particularly smaller ones, while striving to build a more open and compatible AI ecosystem.

\item \textbf{Determining Agent Selection}: 
To tackle the challenge posed by the rapid expansion of AI agents and the increasingly blurred functional boundaries, we propose a context-aware state machine for state transitions, integrating a token logits masking strategy to enable precise and efficient agent selection and routing.

\item \textbf{System Deployment and Validation}: 
We implement the MoMA routing model on a real-world platform and conduct extensive validation. 
Experimental results demonstrate that, compared with existing methods, MoMA not only achieves significant cost savings while maintaining performance comparable to optimal models, but also attains the highest performance under fixed cost constraints.

\end{itemize}


\section{Related Work}

\subsection{Multiple LLMs System}
Most LLM routing aims to assign each incoming query to the LLM most capable of handling it.
P2L \cite{P2L} trains an LLM that takes a natural language prompt as input and outputs a Bradley–Terry \citep{bradley1952rank} coefficient vector to predict human preference votes. The resulting prompt-specific ranking can then be used to guide optimal model routing.
Some existing studies focus on improving routing accuracy or performance. 
ZOOTER \citep{Zooter} introduces a reward-driven routing strategy enhanced by label-based augmentation, aiming to stabilize training and improve reliability.
RouterDC \citep{chen2024routerdc} presents a dual-contrastive learning approach to query routing, which integrates an encoder with LLM-derived embeddings and optimizes through two contrastive objectives to achieve higher routing accuracy.
EmbedLLM \cite{EmbedLLM} utilizes compact learned representations of both queries and models to estimate routing correctness more efficiently.
LLM Blender \citep{llmblender} adopts pairwise model comparisons to identify the top-$k$ candidates for each query and aggregates their outputs to improve overall performance.

Several studies have also explored routing strategies that strike a balance between performance and cost.
RouteLLM \citep{ong2024routellm} trains a binary classifier on preference data to dynamically route queries during inference, selecting between stronger and weaker LLMs. 
AvengersPro \citep{zhang2025beyond}, building on Avengers \citep{zhang2025avengers}, embeds and clusters incoming queries, and then routes them to the most suitable model based on a performance–efficiency score. 
Graph Router \citep{feng2024graphrouter} constructs a heterogeneous graph comprising tasks, queries, and LLM nodes, and leverages edge prediction to estimate performance–cost scores.
Hybrid Router \citep{ding2024hybrid} trains a binary routing function to decide whether a query should be handled by a small or a large LLM. While it achieves a balance between cost and performance, it is limited to only two models, which falls short of the diverse requirements in real-world applications. 
Compared with the above methods, our proposed MoMA router incorporates models with varying parameter scales and trains a powerful router to identify the performance-cost efficient LLM for each user query. This design provides stronger adaptability and compatibility across diverse scenarios. 


\subsection{AI Agents Selection}
In multi-agent systems, agent selection denotes the task of deciding which specialized agent(s) should process a given user input. As LLM-driven applications increasingly integrate dozens of agents, an incorrect selection can cascade through the workflow, triggering unsuitable agent calls, producing unreliable responses. 

Research on agent selection has advanced along three main directions.
Rule-based approaches \citep{rule-fsm1, rule-fsm2} employ predefined heuristics such as keyword matching or pattern recognition to route queries. Although simple and efficient, they lack adaptability and perform poorly when confronted with diverse or unforeseen inputs.
Machine learning approaches \citep{ml-fsm} provide greater flexibility by training classifiers on routing datasets to map user intents to the appropriate agents. However, their effectiveness hinges on access to large, high-quality training data.
LLM-based approaches \citep{LLM-fsm1, LLM-fsm2, LLM-fsm3} now dominate the field. By leveraging the linguistic and reasoning capabilities of LLMs,  enhanced with prompt design, fine-tuning, or retrieval-augmented generation (RAG) \citep{rag}, these methods can assign queries to relevant agents with far greater accuracy. Owing to their adaptability and strong empirical performance, LLM-based routing has become the cornerstone of contemporary multi-agent frameworks.
Nonetheless, existing LLM-based techniques still struggle with precise and reliable selection in large-scale agent repositories, leaving ample room for improvement.

\section{The Framework of MoMA Routing Model}

\begin{figure}[ht]
  \centering
  \includegraphics[width=13.8cm]{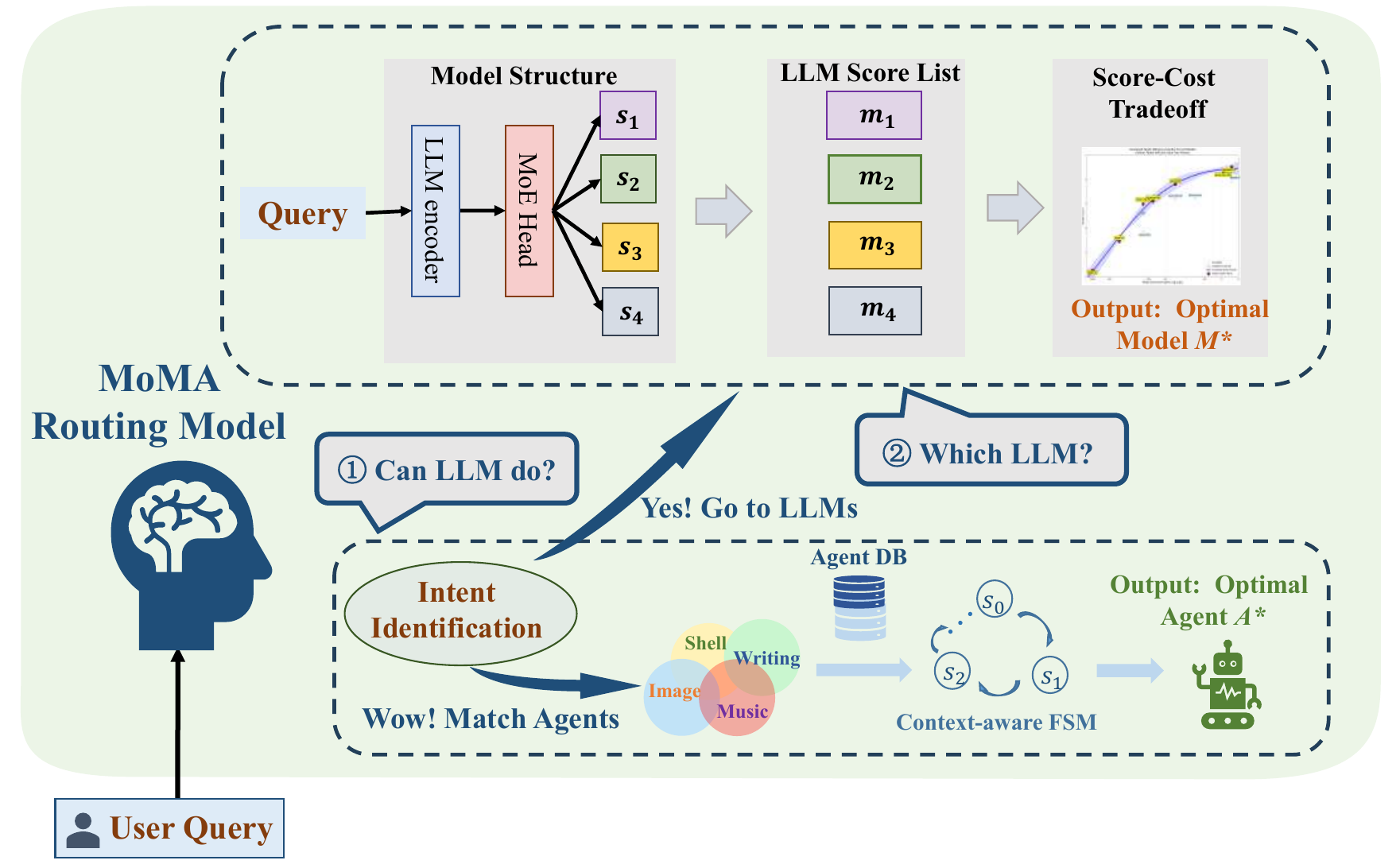}
  \caption{The MoMA routing model framework.}
  \label{framework}
\end{figure}

The overall framework of MoMA routing model is illustrated in Figure~\ref{framework}.
Upon receiving a user query, the trained routing model performs intent recognition to prioritize handling by the agent. Considering the high determinism and enhanced capabilities of task-specific agents, direct matching of the current user query to these agents enables faster and more accurate responses. However, due to the limited number of agents and their functionalities, which cannot cover all user tasks, the routing model will fall back to invoking the LLM when a user request cannot be fulfilled.

\textbf{Can LLM do? (Agent Routing):}
If the current user request can be prioritized for agent handling, the routing model will further select the most appropriate agent.
Inspired by the divide-and-conquer idea, agents are clustered according to their functionalities and descriptions first.  
Then, a context-aware finite state machine is employed for further selection. 
Token logits corresponding to non-selected agents are masked, ensuring that the final choice is made within the correct candidate set. 
This strategy effectively improves routing accuracy without incurring additional cost, particularly in scenarios where the number of agents skyrockets and their functional boundaries become increasingly blurred.

\textbf{Which LLM? (LLM Routing):}
If the user query is assigned to LLM execution, the routing model dispatches the query to the most suitable LLM. 
We explored and validated the performance of different model structures across various task categories and difficulty levels, ultimately confirming the superiority of our proposed routing model structure. 
It estimates the performance score of candidate LLMs based on the rich and augmenting training dataset. 
Based on these predictions, a performance–cost Pareto frontier is constructed. By adjusting weighting factors, our routing model adaptively schedules the performance–cost optimal LLM to respond to the user.

In conclusion, the MoMA routing framework achieves adaptive query routing by first determining whether an LLM should process the user query and then selecting the optimal LLM.
By prioritizing validated and high-efficiency agents, the router avoids the unnecessary cost of invoking expensive models.
During the LLM routing process, the router dynamically explores and selects LLMs with varying parameter sizes based on the specific task requirements, which not only helps small models realise their performance potential but also further reduces the usability overhead for users. 
More importantly, this flexible routing strategy not only improves the efficiency of task execution but also contributes to the development of a more open and compatible AI ecosystem.


\section{Methodology}
\label{Methodology}

\subsection{LLM Routing }

\textbf{Problem Formulation.}
The LLMs in MoMA are denoted as $m \in \mathcal{M} = \{1, \dots, M\}$ with $M$ LLMs, and $\mathcal{D}_{train}$ represents the training dataset. 
The goal is to learn a router that automatically directs each user query to the most appropriate LLM, thereby optimizing both effectiveness and efficiency. 
Formally, given a query $q_i$ as input, the router produces an $M$-dimensional output vector $\bm{r}(q_i) = \big( r_1(q_i), r_2(q_i), \dots, r_M(q_i) \big)$, 
where each component $r_k(q_i)$ reflects the predicted performance score of the corresponding LLM $m_k$ on the given query. 
This vector serves as the basis for selecting the most appropriate LLM to handle the query. 
By further incorporating the cost associated with each LLM, we construct a performance–cost tradeoff curve based on the Pareto frontier, which enables the system to recommend the optimal LLM to different user queries.


\subsubsection{Training Data Construction}
LLMs exhibit varying performance across datasets with different domain coverage, task complexities, and other factors. This diversity places stringent requirements on the datasets used for evaluating LLM capability. 
Consequently, constructing a representative and high-quality training corpus becomes a critical challenge for both model development and performance assessment.

To this end, we constructed a large-scale corpus $D_{train}$, containing approximately \textbf{2.25 million instances}. The corpus is designed to ensure diversity at scale and is systematically partitioned into multiple domains, such as science, writing, technology, and programming, thereby capturing a wide range of real-world application scenarios.
During dataset construction, we emphasized data quality, domain coverage, task diversity, and difficulty levels. Specifically, the corpus was sourced from both open-access and licensed professional texts, followed by systematic cleaning to ensure reliability. 
Each domain distributions were maintained with diverse task types to enhance representativeness. 
The dataset further incorporates multiple task formats alongside a hierarchical design of complexity, from simple to complex, to strengthen generalization. 
Figure~\ref{datadistri} illustrates the distribution of the constructed training dataset. 
In the \textbf{Appendix~\ref{APPcode}}, we provide a detailed analysis of its subcategories in Figure~\ref{code1} and Figure~\ref{code2} using the technology domain as an example, explaining the construction of the enriched dataset and its significance on routing methods.
Overall, $D_{train}$ achieves strong representativeness in terms of \textbf{complexity, domain coverage, and task scale}, providing a solid foundation for model training and evaluation.
The construction of $D_{train}$ not only supplies large-scale, high-quality training samples, but also establishes a unified and reliable platform for performance evaluation and comparative experiments. 

\begin{figure}[ht]
  \centering
  \includegraphics[width=12cm]{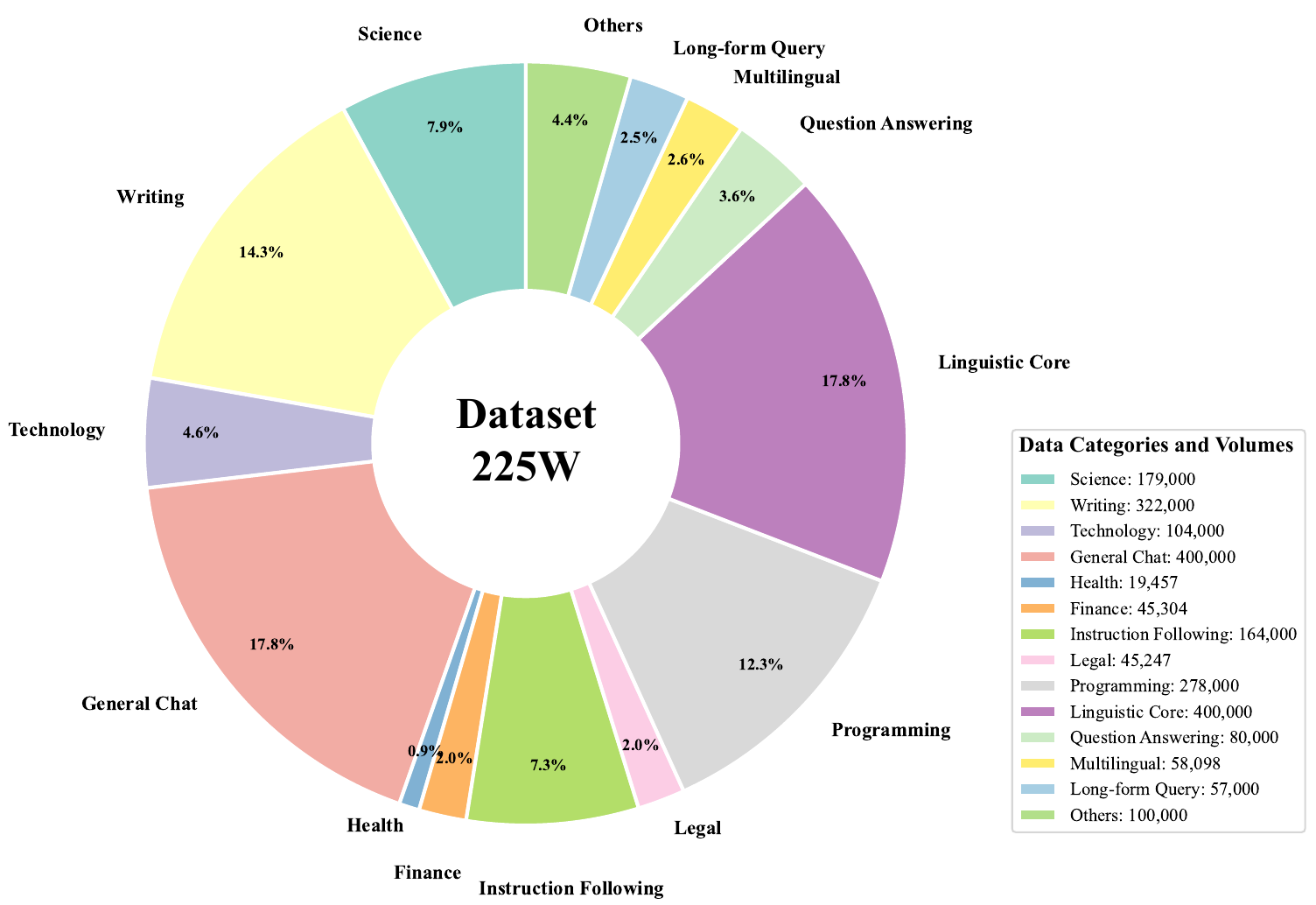}
  \caption{Training data distribution by category.}
  \label{datadistri}
\end{figure}

\subsubsection{Data Augmentation}
To ensure both diversity and representativeness, a BERT-based \citep{devlin2019bert} modeling approach is first employed to select representative query samples from each domain.
Based on these samples, we then design pairwise model comparison tasks and collect the corresponding combating results. 
For evaluation, the LLM-as-a-judge framework is adopted to determine the relative performance of model pairs, resulting in the construction of quadruples in the format $D_i = [q_i, m_a, m_b, \bm{w_i}]$ for each query $q_i$, and $m_a$ and $m_b$ denote LLMs $a$ and $b$, respectively. 
Here, $\bm{w_i}$ characterizes the relative performance between two LLMs under the user query $q_i$, including five possible cases, and we denote $y_k \in \{0, 1, 2, 3, 4\}$ as the probability of these possible scenarios as follows: 
\begin{itemize}
\item $y_k = 0$ corresponds to $m_a = m_b$: the two LLMs perform comparably.
\item $y_k = 1$ corresponds to $m_a > m_b$: LLM $a$ outperforms LLM $b$.
\item $y_k = 2$ corresponds to $m_a < m_b$: LLM $b$ outperforms LLM $a$.
\item $y_k = 3$ corresponds to $m_a \gg m_b$: LLM $a$ significantly outperforms LLM $b$.
\item $y_k = 4$ corresponds to $m_a \ll m_b$: LLM $b$ significantly outperforms LLM $a$.
\end{itemize}

Furthermore, we utilize the Elo rating to establish a quantitative ranking of LLM performance. 

\subsubsection{Router Design}
\begin{figure}[ht]
  \centering
  \includegraphics[width=12cm]{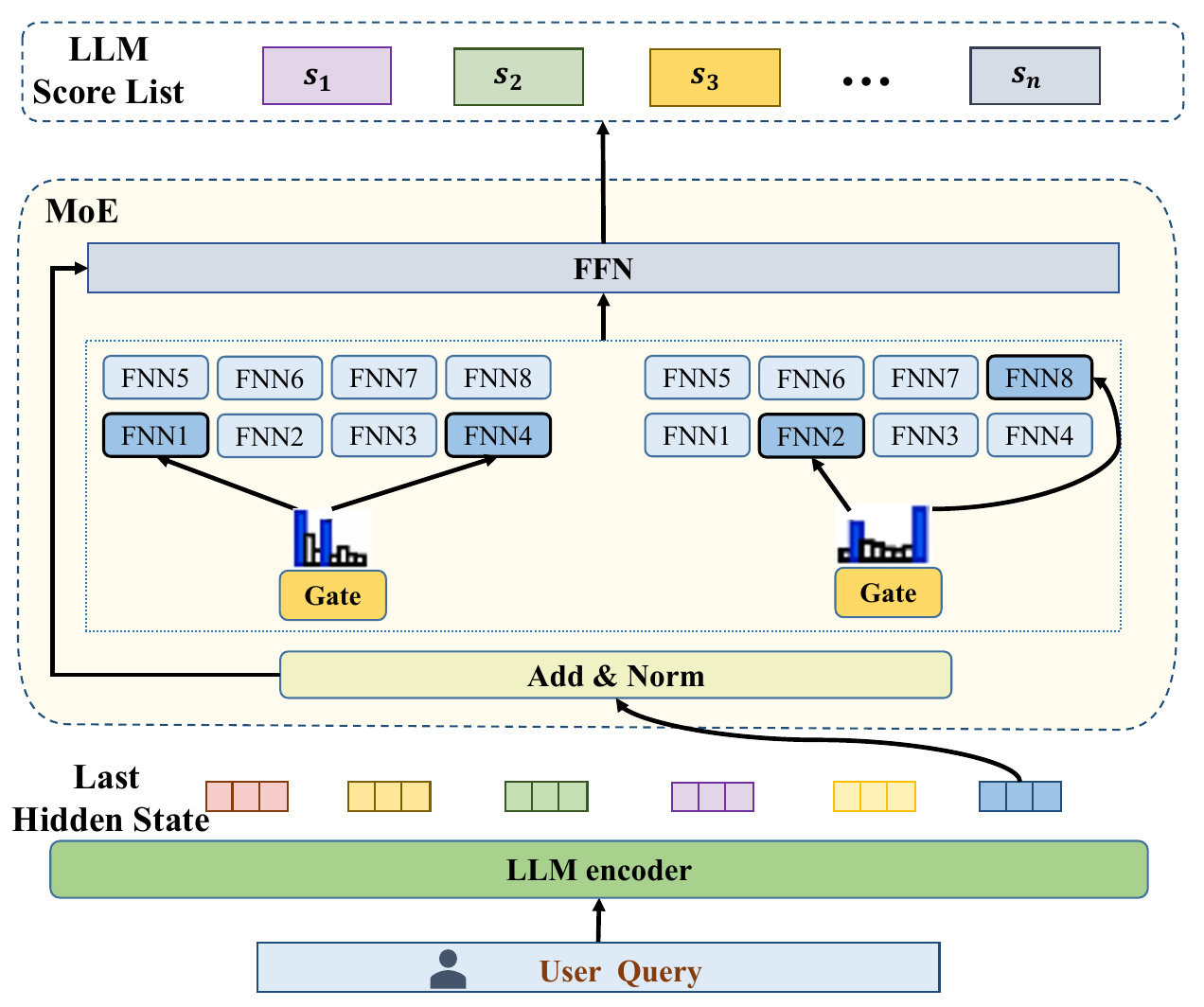}
  \caption{LLM routing network structure.}
  \label{Router1}
\end{figure}

The whole network structure of the multi-LLM router is shown in Figure~\ref {Router1}. 
The user query is fed into the pre-trained instruction-tuned LLM (we use Qwen-3 \citep{qwen3} ) for encoding, and the hidden states of the LLM’s last layer are extracted as feature representations. 
These features are then input into the MOE model head, where a gating network dynamically selects the top-$k$ most suitable experts to process each input. 
The outputs of the activated experts are weighted and summed via the MOE coefficient head to produce the router’s final output, i.e., an $M$-dimensional vector $\bm{r}(q_i)$. Each element of  $\bm{r}(q_i)$ corresponds to the response performance of a specific model based on the current user query $q_i$.

{For user query $q_i$ and LLM pair $[m_a, m_b]$, the outputs of the MOE head are $ [\beta_a, \beta_b ]$ to represent the score of the winner and loser model.} 
For fine-grained prediction of adversarial outcomes, we model the probability distribution over these three outcomes and optimize the model by minimizing the discrepancy between predicted probabilities and ground-truth labels.
$m_a$ outperforms $m_b$ means $m_a > m_b$ and $m_a \gg m_b$, thus we can obtain the three fundamental probabilities ( $m_a$ outperforms $m_b$,  $m_b$ outperforms $m_a$,  $m_a = m_b$) as follows:
\begin{equation}
g_{\theta^*(q_i)}(y_k) = 
\begin{cases} 
\frac{\varphi_a}{\varphi_a + \theta \varphi_b } &  \text{both} \  y_k = 1 \  \text{and} \  y_k=3, \\
\frac{\varphi_b}{\varphi_b + \theta \varphi_a } & \text{both} \ y_k = 2 \  \text{and} \  y_k=4, \\
1 - \frac{\varphi_a}{\varphi_a + \theta \varphi_b } - \frac{\varphi_b}{\varphi_b + \theta \varphi_a }   & y_k = 0,
\end{cases}
\end{equation}
where $\varphi_a = e^{\beta_a}$ and $\varphi_b = e^{\beta_b}$ to ensure that the obtained probability is greater than zero. $\theta \in \mathbb{R}^{N \times 1}$ is a dynamic threshold ( ensuring $\theta > 1$).




Then, to further refine probabilities of winning and losing into ``strong" and ``weak" variants, we use $\delta = \log(\varphi_{\text{win}}) - (\log(\theta) + \log(\varphi_{\text{lose}}))$ to denote the logarithmic advantage of the winner over the adjusted loser: 
\begin{equation}
s_{\text{win}} = \sigma(\kappa (\delta - m)),
\end{equation}
\begin{equation}
s_{\text{lose}} = \sigma(\kappa (-\delta - m)),
\end{equation}
where \( \sigma(x) \) is the sigmoid function, and \( \kappa \) and \( m \) denote comparison strength hyperparameter and margin hyperparameter.
Then the probability of a strong winner and a strong loser can be denoted as:
\begin{equation}
g_{\theta^*(q_i)}(y_k) = 
\begin{cases} 
\frac{\varphi_a}{\varphi_a + \theta \varphi_b } \cdot s_{\text{win}} &   y_k=3, \\
\frac{\varphi_b}{\varphi_b + \theta \varphi_a } \cdot s_{\text{lose} }&   y_k=4. \\
\end{cases}
\end{equation}
Based on the obtained strong winning and losing probabilities, the probabilities of $m_a > m_b$ and $m_a < m_b$ can also be calculated.


The loss function is designed to minimize the discrepancy between the model’s predicted probabilities and the true labels.
Here, we adopt categorical cross-entropy (CCE) to handle the multiple classification task.
True result labels \(Y_i\) (for the \(i\)-th training sample) are converted to \textit{one-hot encoding} to match the 3-class probability output. The loss function \(\mathcal{L}_{\text{GRK}}(\theta^*)\) is defined as:
\begin{equation}
\mathcal{L}_{\text{GRK}}(\theta^*) = -\frac{1}{N} \sum_{i=1}^N \sum_{y_k \in \{0,1,2,3,4\}}  Y_{i} \cdot \log\left( g_{\theta^*(q_i)}(y_k) \right),
\end{equation}
where \(g_{\theta^*(q_i)}(y_k)\) is the model’s predicted probability of the \(i\)-th sample belonging to category \(y_k\), and the negative logarithm \(-\log(\cdot)\) penalizes large deviations between predicted probabilities and true labels. 

The goal of training is to find the optimal parameter function \(\hat{\theta}^*\) that minimizes the categorical cross-entropy loss. Formally, the optimization problem is:
\begin{equation}
\hat{\theta}^* = \underset{\theta^* \in \Theta^*}{\text{argmin}}  \mathcal{L}_{\text{GRK}}(\theta^*),
\end{equation}
where \(\Theta^*\) denotes the space of valid parameter functions mapping prompts to parameter vectors. 

\subsubsection{Score-Cost Tradeoff}

Given a user query $q_i$, we construct a Pareto frontier $\mathcal{M}_i^p$ ( $\mathcal{M}_i^p \in \mathcal{M}$ ) to balance the costs of the LLMs and performance scores output by our routing model, ensuring that the candidate solutions are efficient and cannot be dominated.
The Pareto fronts for user query  $q_i$ can be denoted as:  
\begin{equation}
\mathcal{M}_{i}^{p} = \{ (m_i^k, c_i^k, s_i^k) \mid k=1,\dots,M \},
\end{equation}  
where \(m_i^k\) represents the model name, \(c_i^k \in \mathbb{R}^+\) denotes the inference cost, and \(s_i^k \in \mathbb{R}\) denotes the performance score for user query $q_i$.

By analyzing the Pareto frontier, we utilize the \textbf{TOPSIS} \citep{TOPSIS} algorithm (Technique for Order Preference by Similarity to Ideal Solution) to identify the optimal solution that best satisfies the tradeoff between performance and cost, enabling an efficient and effective model selection.
Firstly, to eliminate scale differences and dimensional inconsistencies, both cost and score are normalized as follows:  
\begin{equation}
c{_i^{k}}'  = \frac{c_i^k - c_{i,min}^k}{ c_{i,max}^k - c_{i,min}^k}, 
\quad 
s{_i^{k}}' = \frac{s_i^k - s_{i,min}^k}{  s_{i,max}^k - s_{i,min}^k},
\end{equation} 
where $c{_i^{k}}'$ and $s{_i^{k}}'$ denote the normalized cost and score. $c{_i^{k}}'$ is expected to be as small as possible, while the $s{_i^{k}}'$ is expected to be as large as possible.  
Then, the ideal point can be denoted as $P^+ = (0,1)$ corresponding to the lowest cost and the highest performance, and the anti-ideal point is $P^- = (1,0)$.

Given the weights $w_c$ and $w_s$ for cost and performance, respectively, the distances of model $m_i^k$ to the ideal and anti-ideal points are computed as
\begin{equation}
d{_i^k }^+ = \sqrt{ \big( w_c \, c{_i^{k}}' \big)^2 + \big( w_s \, (1 - s{_i^{k}}') \big)^2 },
\qquad
d{_i^k }^- = \sqrt{ \big( w_c \, (1 - c{_i^{k}}') \big)^2 + \big( w_s \, s{_i^{k}}' \big)^2 }.
\end{equation}
The relative closeness of $m_i^k$ is then defined as
\begin{equation}
\phi_i^k = \frac{d{_i^k }^-}{d{_i^k }^+ + d{_i^k }^-},
\end{equation}
where a larger $\phi_i^k$ indicates a more desirable tradeoff between performance and cost. 
Finally, we select the LLM with
\begin{equation}
m{_i^k}^\ast = \arg\max_{m_i^k \in \mathcal{M}_i^p} \phi_i^k,
\end{equation}
with ties broken by preferring higher original scores $s_i^k$, and subsequently lower original costs $c_i^k$. 
This procedure ensures that the selected model achieves a balanced compromise between performance and cost, while remaining robust to scale differences and tie cases.

\subsection{Agent Routing }
The design of agent routing follows a divide-and-conquer hierarchical retrieval strategy, which reduces context overhead while improving routing accuracy. 
In the first layer, a coarse-grained classification is performed by grouping agents into high-level categories (e.g., Image, Travel, Meeting). 
It embeds user queries and category descriptions, outputting the top-$k$ most similar categories. 
Subsequently, the second layer utilizes a context-aware state machine to perform fine-grained routing based on the predicted category's output, inspired by context-engineering for AI agents lessons from building Manus by \cite{Manus}. It dynamically loads detailed descriptions of candidate agents under the corresponding category into the LLM's context as needed, completing precise routing.
We will focus on explaining the design of the masking strategy, while the detailed design of the other parts of the algorithm can be found in the \textbf{Appendix~\ref{APPAGENT}}.

\textbf{Token Logits Masking:} The availability of agents is determined by the finite state machine. For unavailable agents, their corresponding token logits are masked during the decoding process to prevent the model from attempting to invoke non-existent or inactive agents. The mask is dynamically generated based on real-time agent status and contextual information, ensuring both the flexibility and robustness of the routing mechanism.  Specifically, during decoding, the model computes the logits distribution for the next possible token. The LLM constructs a mask vector with the same size as the vocabulary and sets the positions corresponding to unavailable agents to $-\infty$. After applying the softmax normalization, the probabilities of these positions are effectively reduced to zero, completely preventing the generation of invalid tokens. By this strategy, the LLM can only select tokens corresponding to valid agent names during inference, thereby ensuring the correctness and safety of agent invocation.

\section{Experiments}
We conducted a comprehensive series of experiments. In this section, we first present a detailed exploration of model architectures, followed by an extensive evaluation of router performance from multiple perspectives to validate its effectiveness.


\subsection{LLM Router Architecture Exploration}
In addition to our proposed routing method based on LLM-as-a-judge combined with a mixture-of-experts architecture, we also explore two alternative routing paradigms: SFT-based classification routing and contrastive learning-based routing.
The detailed introductions are provided in the \textbf{Appendix~\ref{APPSFT}} and \textbf{~\ref{APPCont}}.

The SFT-based approach formulates routing as a multi-class classification task, where the router directly predicts the most suitable model for each prompt. This design is efficient in training and inference but heavily depends on the availability of fine-grained labels and suffers when task boundaries are ambiguous.
In contrast, the contrastive learning-based approach leverages a strong judge model (we use Gemini2.5 \citep{2025gemini}) to generate preference signals. By constructing positive and negative response pairs, the router learns a representation space that captures fine-grained differences between models. This method improves robustness and scalability but requires substantial training cost and large-scale annotations from the judge model.

For clarity, we provide a comparative summary of these three routing approaches across multiple dimensions, as shown in Table~\ref{tab-routing}.
\begin{table}[h]
\centering
\caption{Comparison of three routing approaches across multiple dimensions.}
\label{tab-routing}
\resizebox{\textwidth}{!}{
\begin{tabular}{p{3.2cm} p{3.2cm} p{3.2cm} p{3.2cm}}
\hline
\textbf{Dimension} & \textbf{SFT-based Classification Router} & \textbf{Contrastive Learning-based Router} & \textbf{MoMA Router (Ours)} \\
\hline
Dataset Construction Difficulty & High: requires clear $(x, m^\ast)$ labels & High: requires multiple responses per prompt and judge scoring & Medium: only two responses per prompt with judge evaluation \\
\hline
Sensitivity to Category Boundaries & High: performance drops with fuzzy categories & Low: captures fine-grained differences in continuous space & Medium: mitigated by score-based evaluation \\
\hline
Scalability & Poor: adding new models requires laborious and fine-grained relabeling & Medium: new models can be integrated by retraining and generating prototypes & Medium: supporting retraining to lean model profile \\
\hline
Inference Efficiency & High: single forward classification & Medium: requires similarity computation or prototype comparison & Medium: needs expert scoring and routing \\
\hline
Main Advantages & Simple, interpretable, efficient deployment & Robust, generalizable, flexible extension & Objective evaluation, adaptive routing with MoE \\
\hline
Main Limitations & Strong label dependence, weak generalization & Expensive training, judge bias risks & Moderate cost, dependent on LLM as judge \\
\hline
\end{tabular}}
\end{table}

The three routing strategies exhibit distinct characteristics.
While the SFT-based classification approach is simple and efficient, it relies heavily on well-defined labels and exhibits limited generalization. 
The contrastive learning-based method offers greater flexibility and robustness, but at the expense of high training costs and potential bias from the judge model. 
In comparison, our proposed MoMA router, which integrates LLM-as-a-judge with a MoE architecture, strikes a stronger balance across key criteria: it reduces dependence on extensive labeled data and mitigates challenges from ambiguous task boundaries through score-based evaluation. Furthermore, the inherent flexibility of the MoE structure supports scalable model expansion. 
Our method provides an adaptive and highly scalable routing at a lower cost, offering a more practical and sustainable solution for efficient utilization of heterogeneous models.

\subsection{Performance Comparison}

\subsubsection{Experimental Setting}

\textbf{Benchmarks.} To evaluate the generalization ability of our router across diverse domains, we conducted experiments on several widely adopted public benchmarks.
\begin{itemize} [leftmargin=0pt, itemindent=0pt, labelindent=0pt]
\item \textbf{AIME2024 \citep{AIME}: } A benchmark derived from the American Invitational Mathematics Examination 2024, consisting of complex mathematical problems designed for high-school level competitions. The dataset requires advanced mathematical reasoning, algebraic manipulation, and problem decomposition, serving as a rigorous test of a model’s higher-order problem-solving and generalization abilities. 

\item \textbf{LiveCodeBench \citep{LiveCodeBench}: } A large-scale benchmark for code generation and execution-based evaluation, collected from competitive programming platforms and real-world software repositories. It covers multiple programming languages and problem types, requiring not only syntactic correctness but also semantic precision verified through execution. The benchmark evaluates a model’s ability to generate functional, efficient, and robust code in diverse scenarios. 

\item \textbf{SimpleQA \citep{SimpleQA}: } A lightweight benchmark designed for factoid-style question answering over general knowledge domains. The dataset contains short, single-hop questions that can typically be answered with concise factual information. It serves as a measure of a model’s ability to retrieve, comprehend, and directly respond to straightforward natural language queries with high accuracy. 



\end{itemize}


\textbf{Candidate LLMs.} We compare our router across a diverse set of LLMs with varying parameter scales, including both widely used open-source models and multiple proprietary models developed by China Mobile’s Jiutian series, as shown in Table~\ref{LLMcost}. This selection allows us to assess routing effectiveness under heterogeneous architectures and parameter capacities.
\begin{itemize} [leftmargin=0pt, itemindent=0pt, labelindent=0pt]
\item  \textbf{deepseek-r1 \citep{deepseekr1}:} A reasoning-focused model designed to enhance logical inference and multi-step problem-solving.

\item \textbf{deepseek-v3 \citep{deepseekv3}:} A general-purpose LLM optimized for broad natural language understanding and generation.

\item \textbf{qwen2.5-code-32b \citep{qwen2.5}:} A 32B-parameter code-oriented model from the Qwen series, specialized for program synthesis, debugging, and code completion.

\item \textbf{qwen3-32b \citep{qwen3}:} The third-generation 32B-parameter general-purpose Qwen model, offering improved performance in reasoning and natural language tasks.

\item \textbf{qwen3-235b-a22b \citep{qwen3}:} A large-scale mixture-of-experts model with 235B parameters and 22B activated parameters, designed to balance efficiency and performance across complex tasks.
\end{itemize}

Jiutian series (China Mobile) \footnote{https://jiutian.10086.cn/}:
\begin{itemize} [leftmargin=0pt, itemindent=0pt, labelindent=0pt]
\item \textbf{jiutian-1b:} A lightweight 1B-parameter model tailored for low-latency inference and resource-constrained scenarios.

\item \textbf{jiutian-3b:} A medium-scale model with 3B parameters, providing stronger general-purpose capabilities while maintaining efficiency.

\item \textbf{jiutian-8b:} A general-purpose 8B-parameter model designed for more complex reasoning and generation tasks.

\item \textbf{jiutian-code-8b:} An 8B-parameter code-specialized model optimized for software development and engineering applications.

\item \textbf{jiutian-math-8b:} An 8B-parameter model tailored for mathematical problem solving and quantitative reasoning.

\item \textbf{jiutian-lan-13b: }A 13B-parameter model optimized for language understanding and generation, with enhanced fluency and robustness.

\item \textbf{jiutian-lan-comv3: } An advanced 75B-parameter commercial variant of the Jiutian language model, offering improved accuracy and adaptability across enterprise applications.

\end{itemize}

This comprehensive model set, ranging from lightweight 1B-parameter systems to large-scale MoE architectures, ensures a robust evaluation of our router’s scalability and adaptability across heterogeneous model pools.

\begin{table}[t]
\caption{Candidate LLMs information (Aliyun Bailian).}
\label{LLMcost}
\begin{center}
\begin{tabular}{ccc}
\hline
\textbf{LLM}  & \begin{tabular}[c]{@{}c@{}} \bf{Input {Price}}\\ (\textyen /1K tokens)\end{tabular} & \begin{tabular}[c]{@{}c@{}} \bf{Output}\\ (\textyen/1K tokens)\end{tabular} \\ \hline
{deepseek-r1}      & 0.004              & 0.016              \\
{deepseek-v3}      & 0.004              & 0.012               \\
{qwen2.5-code-32b} & 0.002              & 0.006             \\ 
{qwen3-32b}        & 0.002              & 0.02             \\
{qwen3-235b-a22b}   & 0.002              & 0.02            \\
{jiutian-1b}       & 0.0003            & 0.0012             \\ 
{jiutian-3b}       & 0.0003             & 0.0012              \\
{jiutian-8b}       & 0.0005              & 0.002              \\
{jiutian-code-8b}  & 0.001              & 0.002              \\ 
{jiutian-math-8b}  & 0.001             & 0.002              \\
{jiutian-lan-13b}  & 0.001              & 0.0038              \\
{jiutian-lan-comv3} & 0.004             & 0.012              \\ \hline
\end{tabular}
\end{center}
\end{table}





\subsubsection{Exploring among Different Parameters LLMs}

\begin{figure}[ht]
  \centering
  \includegraphics[width=10cm]{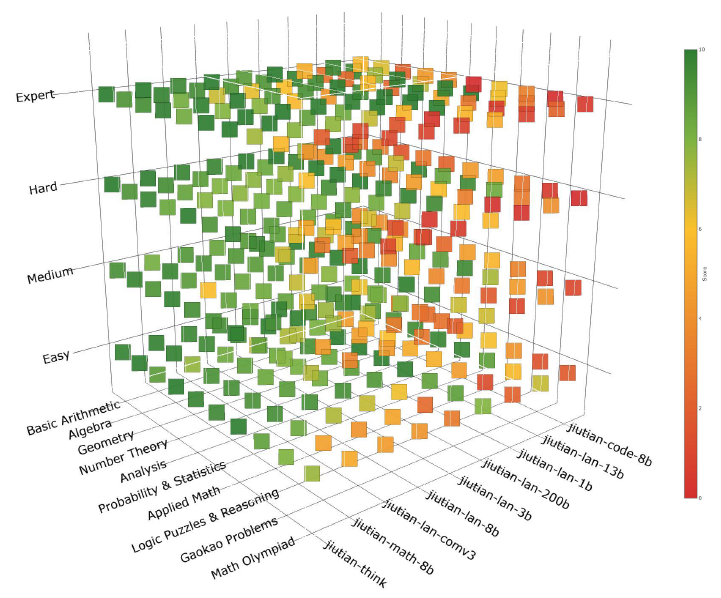}
  \caption{Exploring the performance of Jiutian serial LLMs in the mathematics domain.}
  \label{parameters}
\end{figure}

To validate the representativeness of our proposed method for model capabilities, we conducted experiments and visualized the results.
We illustrate this using the Jiutian series model with varying parameter scales in the field of mathematics as an example, which can bes seen in Figure~\ref{parameters}. 
This three-dimensional heatmap illustrates the performance of various parameter configurations and domain models within the Jiutian series across different mathematical subfields (including elementary arithmetic, algebra, geometry, number theory, etc.) and difficulty levels (from easy to expert-level). The models primarily include jiutian-lan-1b, jiutian-lan-3b, jiutian-lan-8b, jiutian-math-8b, jiutian-code-8b, jiutian-lan-13b, jiutian-lan-comv3 (75b), jiutian-think (75b), and jiutian-lan-200b.

The color gradient, from red (indicating poor performance) to green (indicating excellent performance), quantifies the model performance. It can be observed that most models achieve favorable performance (shown in green) in the Easy difficulty level across all mathematical subfields. However, as the difficulty increases to Medium, Hard, and especially Expert, the performance degrades significantly.
Additionally, distinct models exhibit varying performance patterns in different mathematical subfields at different difficulty levels, further demonstrating the effectiveness of our proposed method in characterizing model capabilities comprehensively.

\subsubsection{MoMA score-cost trade-off}
\begin{figure}[ht]
  \centering
  \includegraphics[width=10cm]{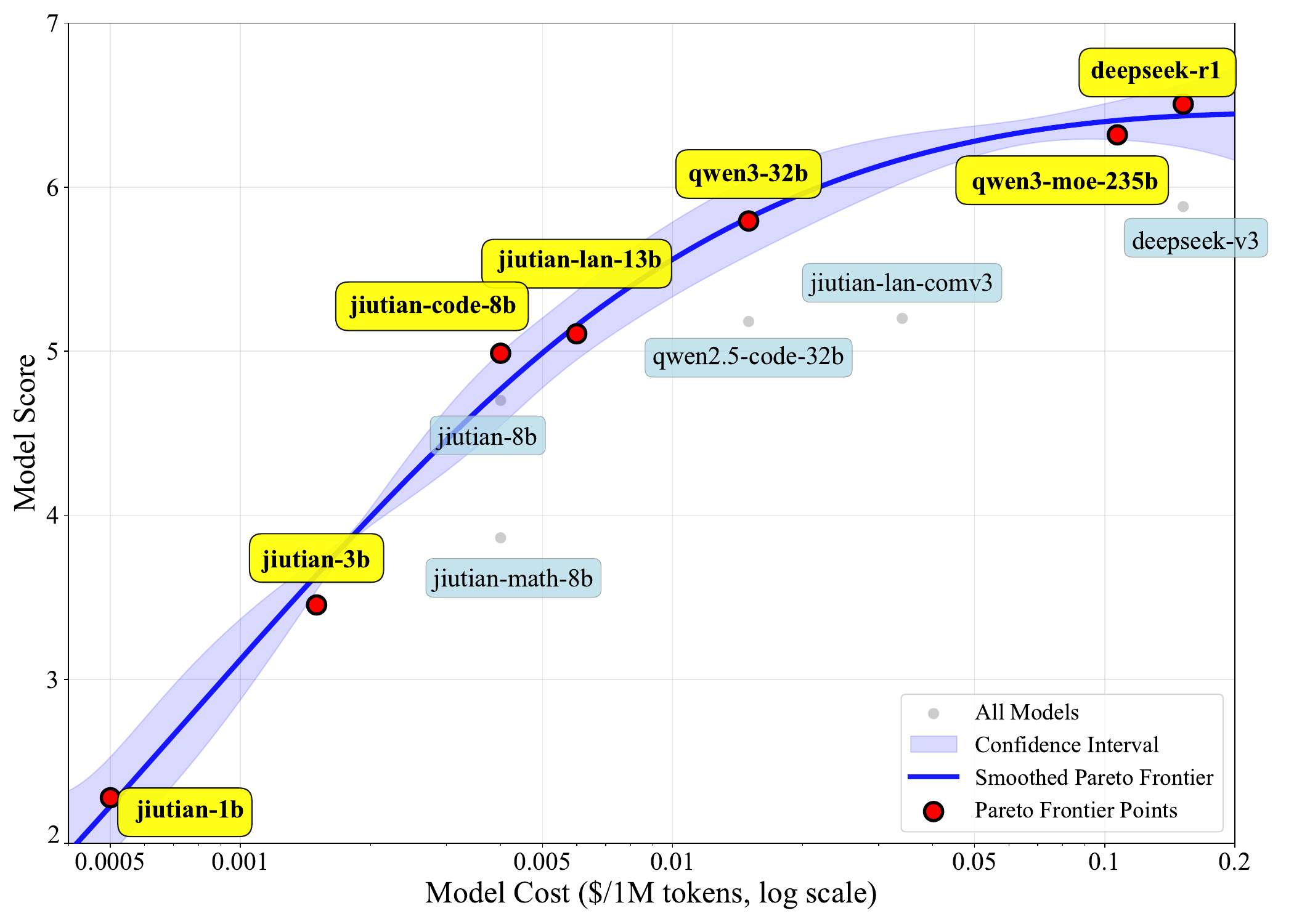}
  \caption{The Pareto frontiers curve for score-cost.}
  \label{score-cost}
\end{figure}

Figure~\ref{score-cost} illustrates the Pareto frontier fitting curves for the input user query.  
During the inference phase, user input queries can be mapped to corresponding task scenarios, facilitating the dynamic routing of the optimal LLM.
After obtaining the scores of each LLM for the current user input based on the routing model, we combine the LLMs' FLOPS to generate a Pareto frontier curve for score-cost using Pareto optimization, as shown in Figure~\ref{score-cost}.
In this figure, the gray points represent all models, with red points indicating Pareto frontier points. The blue line depicts the frontier curve fitted to these points, exhibiting a certain variance.
Building upon this, we integrate the aforementioned TOPSIS algorithm to output the optimal model that best meets user requirements.


Additionally, it is worth noting that MoMA supports dynamic LLM selection based on user preference.
(1) \textbf{Performance-priority}: The model with the best performance is prioritized. (2) \textbf{Cost-Priority}: The optimal solution is selected within the specified cost range. (3) \textbf{Automatic routing}: Both performance and cost are evaluated comprehensively to achieve a dynamically balanced selection.

\subsubsection{Comparison for different routing models}
\textbf{Comparison with single LLM:} When evaluating six single models, qwen3-235b-a22b achieves the highest score (68.6) across three benchmarks. Deepseek-r1 followed closely with 60.2, as shown in Table~\ref{table}.
Compared to a single LLM, MoMA achieves state-of-the-art performance in both AIME2024 and SimpleQA benchmarks under performance-priority scenarios. Compared to the optimal single model (qwen3-235b-a22b), it achieves comparable performance (with a 2.9\% score improvement) while reducing costs by 31.46\%.

\textbf{Comparison with other routing frameworks: } 
MoMA router with the performance-first preference achieves optimal performance. Its automated routing strategy achieves a relatively high score (surpassing deepseek-v3) at a significantly lower cost (37.19\% reduction compared to the performance-priority), thereby achieving an optimal trade-off between performance and cost. 
The SFT-based approach, with only an optimizing model as output, fails to achieve a cost-performance trade-off.
Although it performs best under the auto-routing preference across the three routing frameworks, this advantage stems from our relatively constrained data categories, such methods perform well under limited category conditions. However, in practical applications involving numerous categories, its performance degrades significantly. Moreover, its computational cost is higher than the other two auto-routing frameworks, achieving only marginal performance gains.
Contrastive learning-based methods exhibit performance comparable to MoMA, yet MoMA achieves lower computational and training costs among the three preferences.



\begin{table}[htbp]
  \caption{Performance and cost comparison of MoMA with single-model and other routing methods.}
  \centering
    \label{table}
  \resizebox{\textwidth}{!}
  {
\begin{tabular}{cccccc|c}
\hline
\multicolumn{2}{c}{LLMs}                                                                                                                  & AIME2024      & LiveCodeBench & SimpleQA      & Average Score  & Cost \\ \hline
\multicolumn{2}{c}{deepseek-r1}                                                                                                           & 79.8          & \textbf{73.1} & 27.8          & 60.2          & 12.327    \\
\multicolumn{2}{c}{deepseek-v3}                                                                                                           & 59.4          & 27.2          & 24.9          & 37.2          & 9.498    \\
\multicolumn{2}{c}{qwen3-32b}                                                                                                             & 81.4          & 60.7          & 8.0           & 50.0          & 14.65    \\
\multicolumn{2}{c}{qwen3-235b-a22b}                                                                                                        & 85.7 & 65.9          & 54.3 & 68.6 & 14.65    \\
\multicolumn{2}{c}{jiutian-math-8b}                                                                                                       & 37.5          & -             & -             & -              & 1.667    \\
\multicolumn{2}{c}{jiutian-code-8b}                                                                                                       & -             & 26.3          & -             & -              & 1.667    \\ \hline
\multicolumn{1}{c|}{\multirow{3}{*}{\textbf{MoMA Router}}}                                                         & cost-proirity        & 35.8          & 24.6          & 12.1          & 24.2          & 1.357    \\
\multicolumn{1}{c|}{}                                                                                              & auto-routing         & 65.2          & 45.3          & 19.5          & 43.3          & 6.306    \\
\multicolumn{1}{c|}{}                                                                                              & performance-priority & \textbf{87.3}          & 66.5          & \textbf{56.3}          & \textbf{70.1 }          & 10.04   \\ \hline
\multicolumn{1}{c|}{\begin{tabular}[c]{@{}c@{}}SFT-based \\ Classification Router\end{tabular}}                     & auto-routing         & 76.8          & 70.5          & 40.7          & 62.7          & 8.667    \\ \hline
\multicolumn{1}{c|}{\multirow{3}{*}{\begin{tabular}[c]{@{}c@{}}Contrastive learning \\ based Router\end{tabular}}} & cost-proirity        & 31.7          & 27.6          & 14.2          & 24.5           & 1.667    \\
\multicolumn{1}{c|}{}                                                                                              & auto-routing         & 65.7          & 40.1          & 17.8          & 41.2           & 6.940    \\
\multicolumn{1}{c|}{}                                                                                              & performance-priority & 81.2          & 61.3          & 38.7          & 60.4           & 12.498    \\ \hline
\end{tabular}}
\end{table}



\subsubsection{Distribution of model usage}
\begin{figure}[htbp]
  \centering
  \subfigure[Model usage percentage within the \textbf{code domain} under different preferences. ]{%
    \begin{minipage}{\textwidth}
      \centering
      \includegraphics[width=0.32\textwidth]{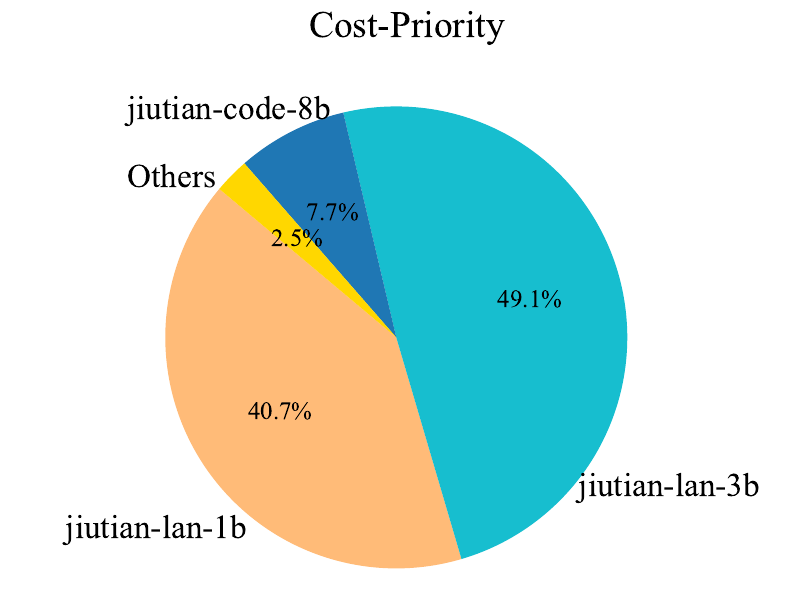}
      \includegraphics[width=0.32\textwidth]{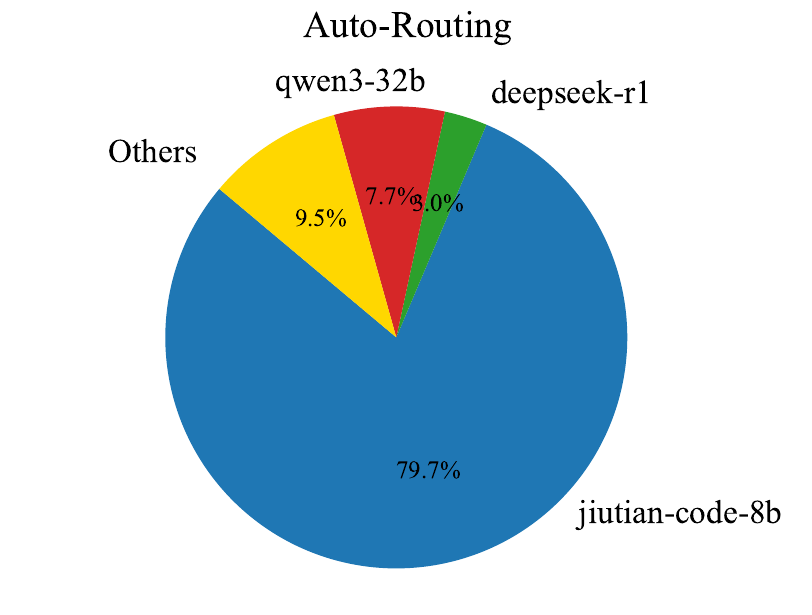}
      \includegraphics[width=0.32\textwidth]{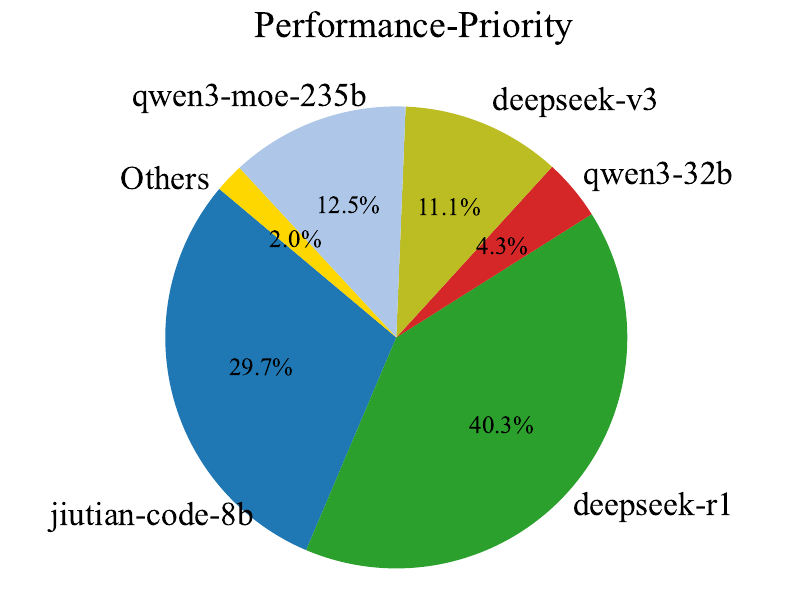}
    \end{minipage}
  }
  \subfigure[Model usage percentage within the \textbf{mathematical domain} under different preferences. ]{%
    \begin{minipage}{\textwidth}
      \centering
      \includegraphics[width=0.32\textwidth]{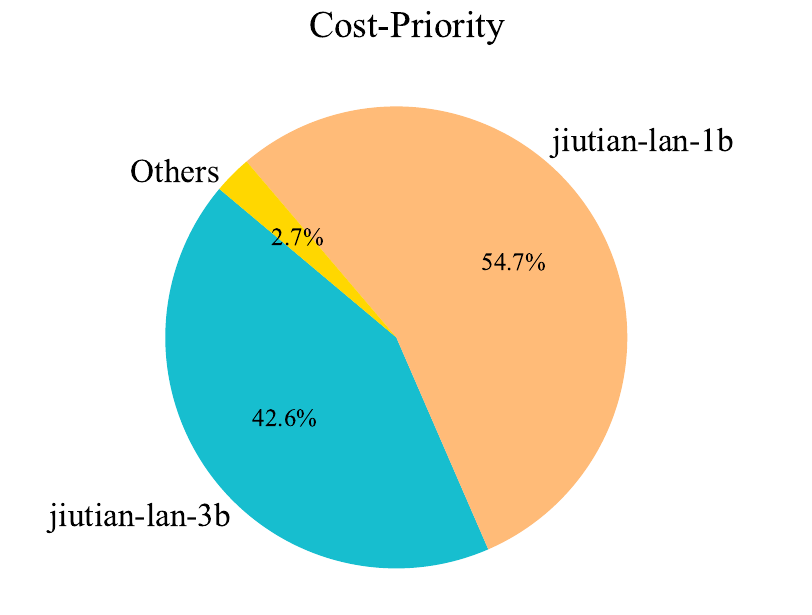}
      \includegraphics[width=0.32\textwidth]{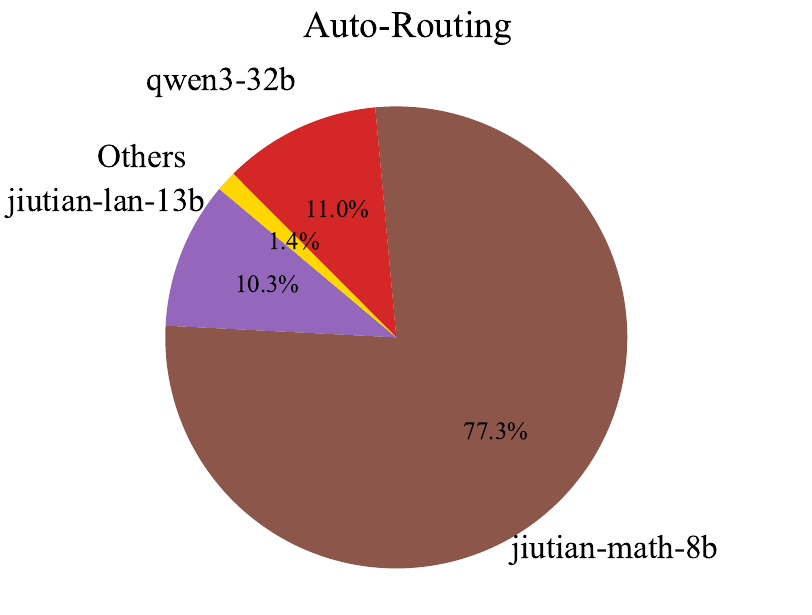}
      \includegraphics[width=0.32\textwidth]{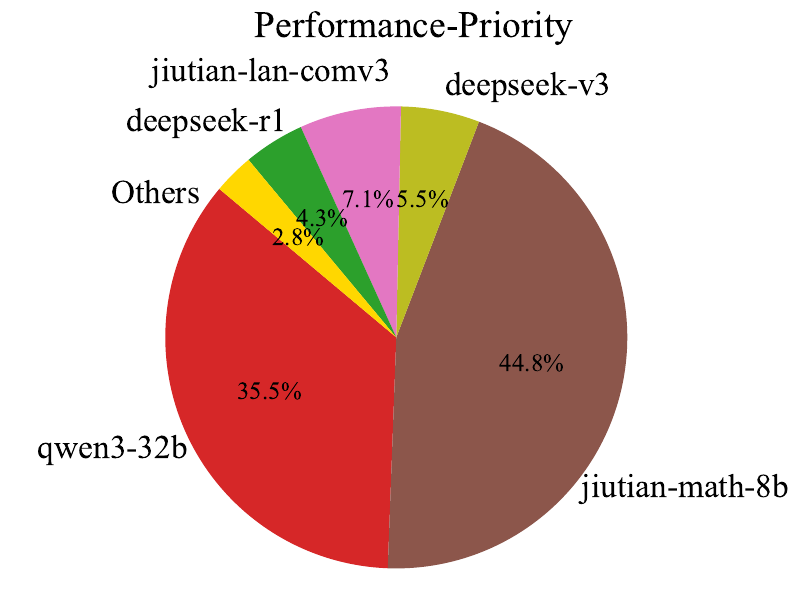}
    \end{minipage}
  }
  \subfigure[Model usage percentage within the \textbf{general knowledge domain} under different preferences.]{%
    \begin{minipage}{\textwidth}
      \centering
      \includegraphics[width=0.32\textwidth]{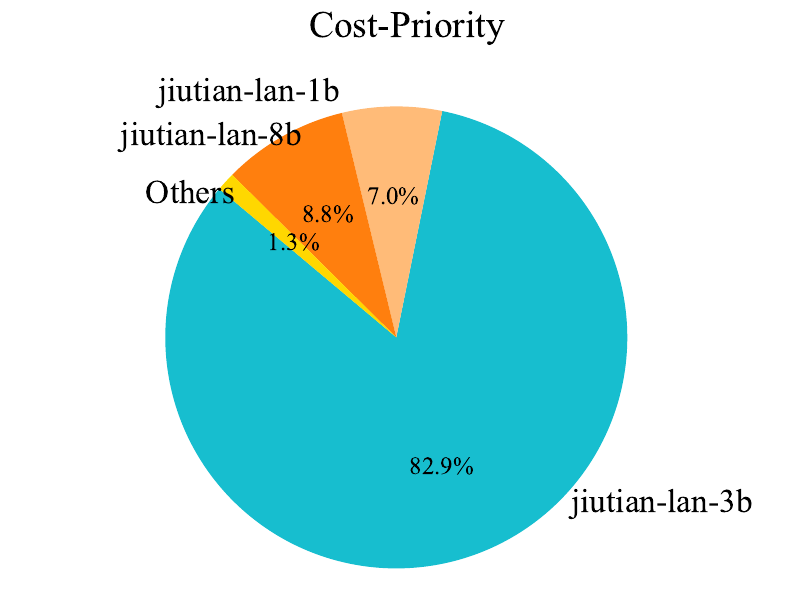}
      \includegraphics[width=0.32\textwidth]{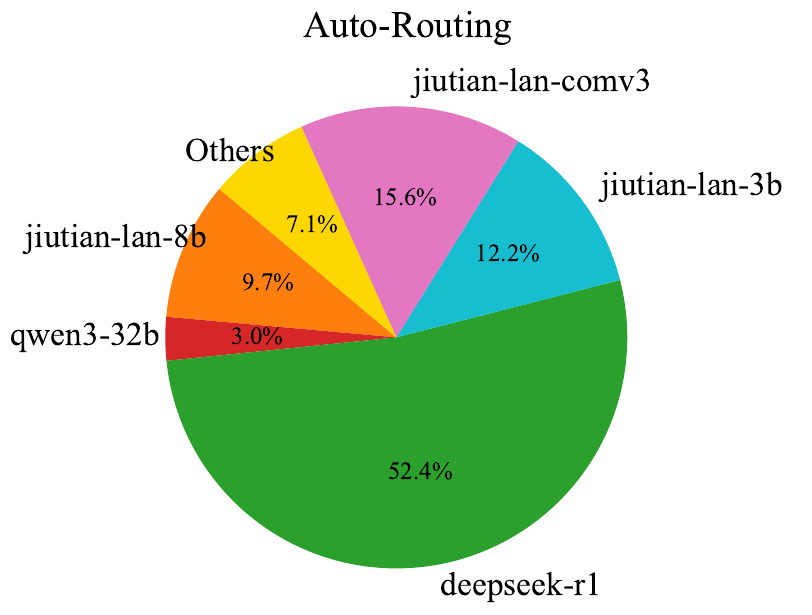}
      \includegraphics[width=0.32\textwidth]{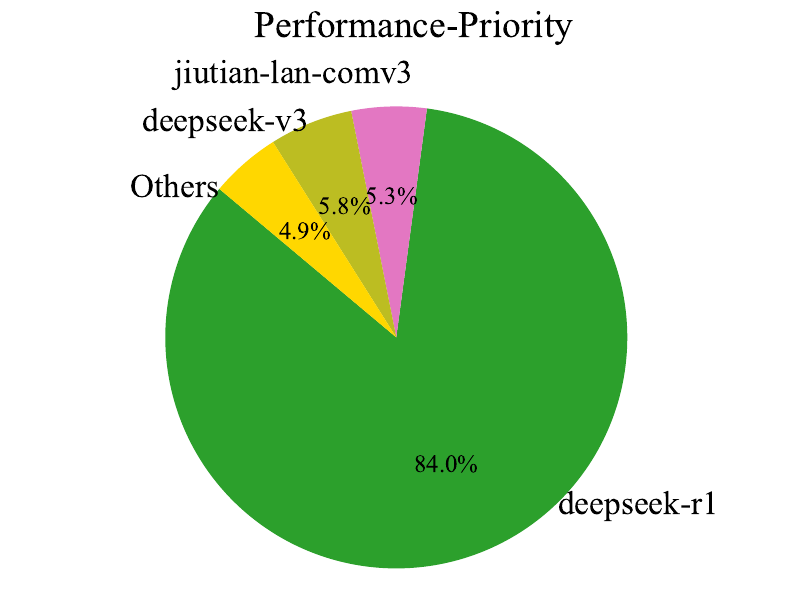}
    \end{minipage}
  }
  \caption{Model usage percentage across code, mathematical, and general knowledge domains. Each domain corresponds to three preferences: cost-priority, auto-routing, and performance-first (from left to right).}
  \label{proportion}
\end{figure}

Figure~\ref{proportion} illustrates the distribution of model usage within the MoMA framework across three benchmark datasets (coding, mathematics, and general knowledge) under different user preference settings (cost-priority, auto-routing, and performance-priority (from left to right)). 

The analysis demonstrates MoMA’s remarkable ability to automatically route and dynamically orchestrate, enabling effective and reliable inference of complex tasks by fully leveraging the strengths of various models.
Under cost-priority preferences, jiutian-lan3b is utilized most extensively across all three benchmarks, with a particularly dominant role in general writing tasks. Under performance-priority preferences, the widely recognized deepseek-r1 is heavily employed in general writing, while the domain-specialized models jiutian-math-8b and jiutian-code-8b excel in mathematics and coding, respectively, thereby ensuring optimal task-specific performance.
In the automatic routing setting, MoMA dynamically invokes models that balance cost and performance across different domains, enabling near-optimal results at a significantly lower cost. For instance, compared to the performance-oriented setting, jiutian-math-8b is adopted more frequently in the mathematics benchmark, offering users strong performance at a reduced cost.

These findings not only highlight the adaptability and effectiveness of the MoMA but also bring attention to the underappreciated role of specialized lightweight models. 
This sheds light on their value in building a more enriched and inclusive AI ecosystem.

\subsection{Real-world Application}
MoMA has been successfully deployed 
with a dozen high-quality models, including the Jiutian, Qwen, DeepSeek, and other series.
They span both general-purpose and specialized domains, covering areas such as programming, mathematics, translation, and healthcare. 
Additionally, over 20 expert agents have been integrated, including tools for daily management, meeting assistants, Migu Music, and deep reporting, all designed to precisely match user requirements and assist users in quickly resolving domain-specific issues.


\section{Conclusion}
To address complex heterogeneous user requests and the growing diversity of capabilities in LLMs and agents, this paper proposes a generalized routing model MoMA that adaptively directs queries to the most appropriate LLM and agent, aiming to achieve an efficient and reliable AI inference for complex task scenarios and an optimized tradeoff of performance and cost. 
We first constructed a large, rich dataset for meticulous classification. Building upon this foundation, we explored and validated three routing frameworks, demonstrating that our proposed MoMA routing framework achieves more practical, scalable, and adaptive routing at a lower cost. 
Experiments across extensive datasets, three routing frameworks, and 12 LLMs demonstrate that MoMA substantially reduces routing costs while maintaining near-optimal model performance, and it delivers state-of-the-art performance at comparable costs when compared to the most strong single LLM.
Thus, the MoMA router strikes an effective balance between performance and overhead, which lays a solid foundation for an economically sustainable future generalized routing frameworks and the AI ecosystem.

\section*{Acknowledgments}

\bibliography{iclr2025_conference}
\bibliographystyle{iclr2025_conference}

\appendix
\section{Appendix}

\subsection{Detailed Training Data Distribution}
\label{APPcode}
Figure~\ref{datadistri} presents the overall distribution of the constructed training data across different domains, with each domain further divided into multi-level subcategories. Such a hierarchical organization not only ensures comprehensive coverage of diverse user tasks but also provides explicit structural signals that guide the routing model in task identification and decision-making at multiple levels of granularity.

\begin{figure}[ht]
  \centering
  \includegraphics[width=9cm]{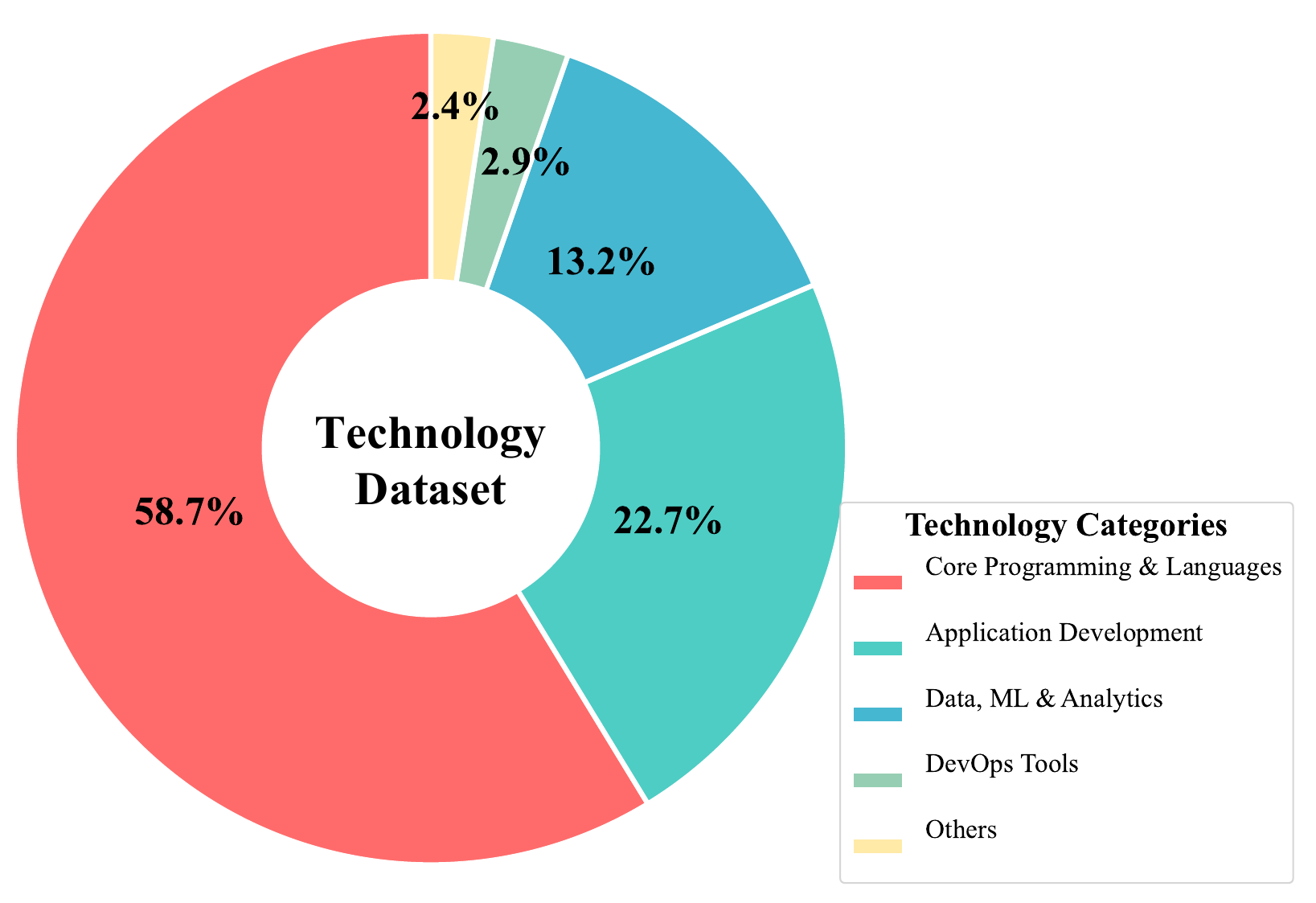}
  \caption{Training data distribution in the technology domain.}
  \label{code1}
\end{figure}

Taking the technology domain as an example, Figure~\ref{code1} illustrates its second-level category distribution. The Core Programming and Languages category accounts for nearly half of the data, occupying a dominant position. Although the distribution appears imbalanced, this reflects the characteristics of real-world tasks: core programming languages and related problems naturally occur with much higher frequency. Thus, the imbalance is not a bias to be corrected, but rather a necessary design choice to ensure that the model acquires sufficient capacity on high-frequency tasks.

\begin{figure}[ht]
  \centering
  \includegraphics[width=14cm]{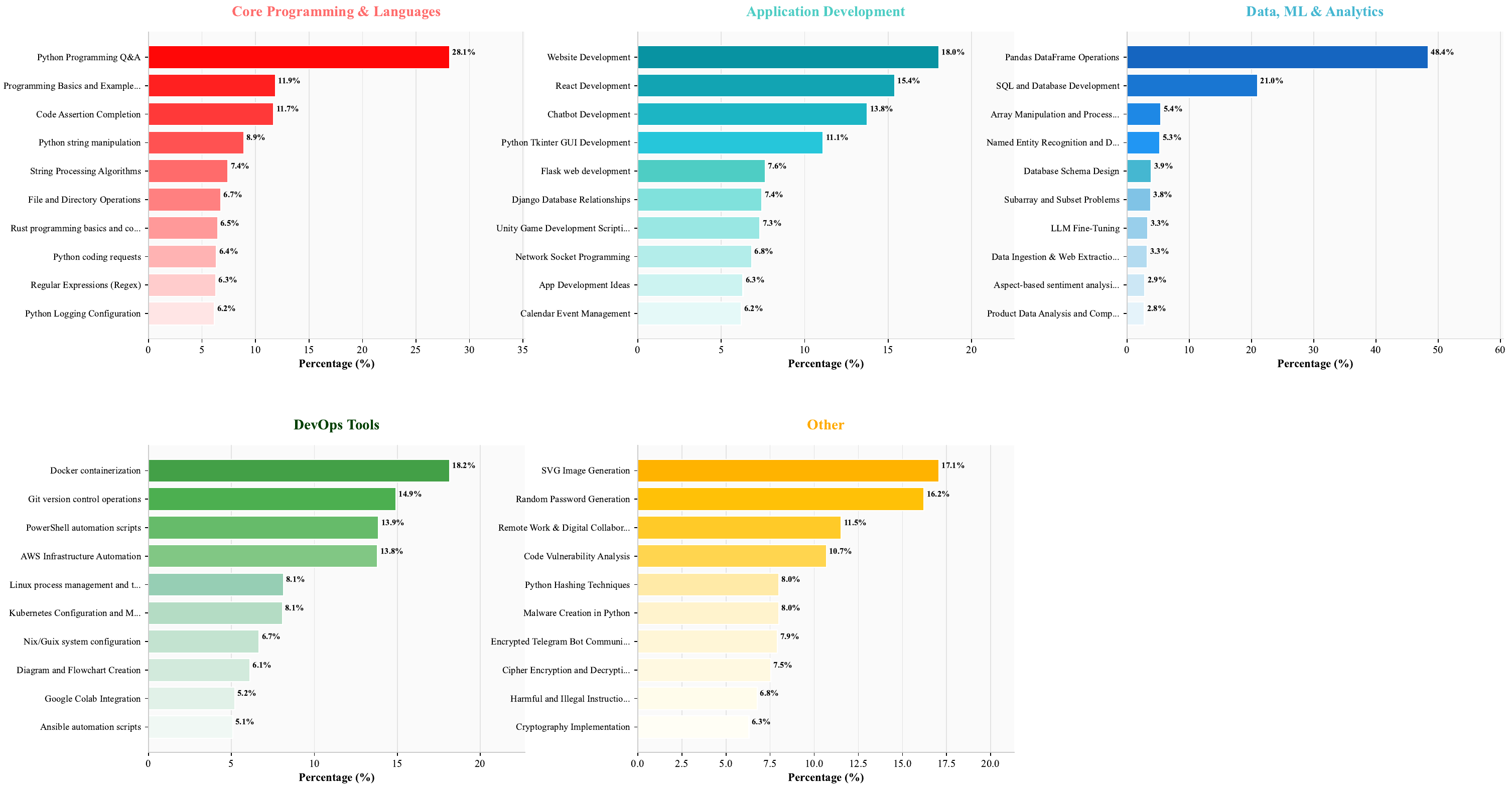}
  \caption{Detailed subcategory data distribution in the technology domain.}
  \label{code2}
\end{figure}

Figure~\ref{code2} further expands the second-level categories in Figure~\ref{code1} into third-level subcategories, revealing a more fine-grained distribution. The relative proportions of these subcategories remain consistent with real-world task patterns. This hierarchical expansion enhances both the authenticity and representativeness of the dataset, while simultaneously enabling the model to leverage both “macro-level category signals” and “fine-grained distinctions” during routing. 
In summary, the hierarchical construction of categories in the technology domain provides more than just a realistic distribution of training data—it also establishes methodological foundations for routing model design.


\subsection{Agent Routing Design}
\label{APPAGENT}
The two-layer design prevents the context window from expanding as the agent pool grows. Since each layer only handles a limited set of candidates, inference can run efficiently and in parallel, enabling scalable routing across large agent collections.

\begin{figure}[ht]
  \centering
  \includegraphics[width=13cm]{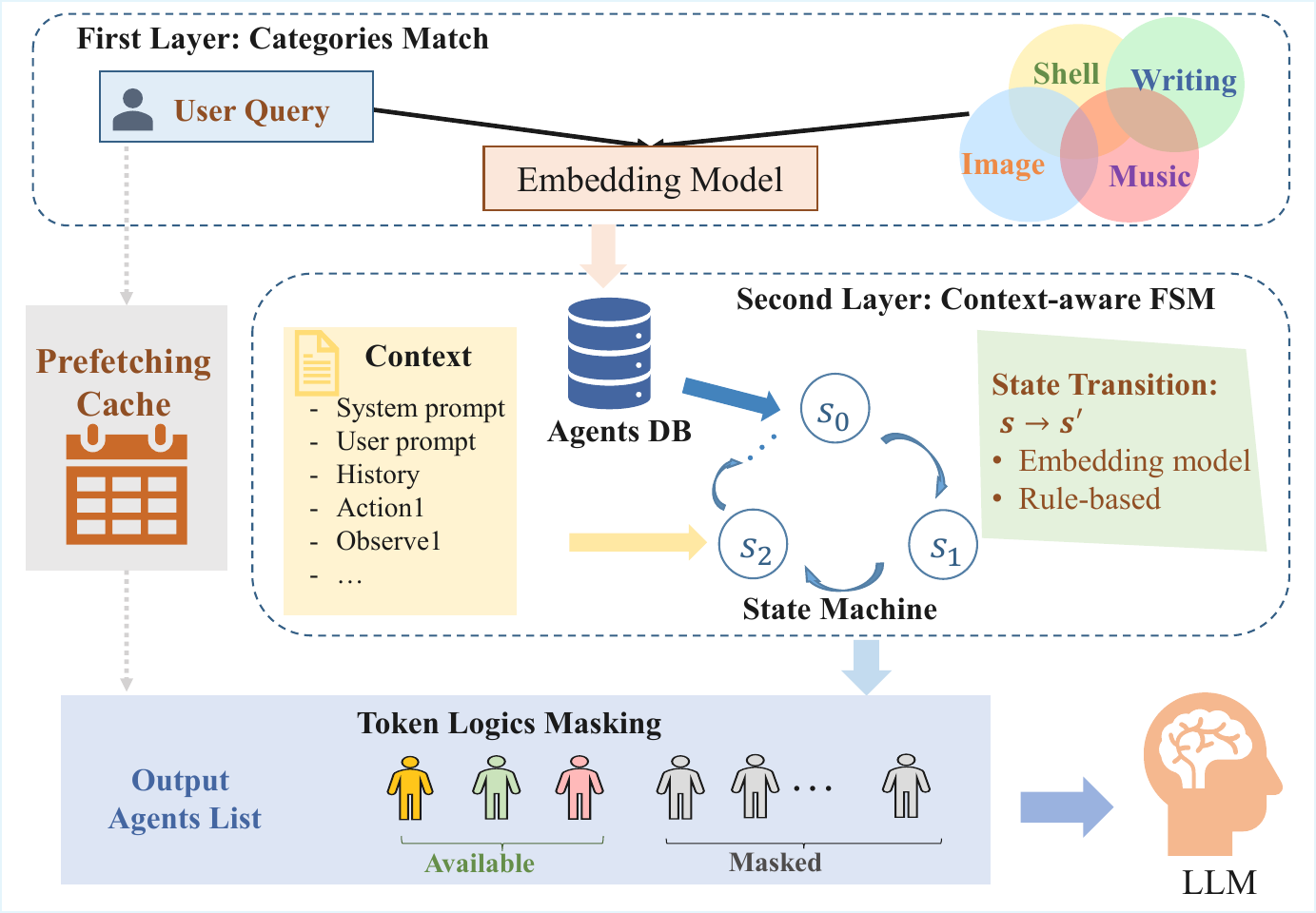}
  \caption{The Agent Routing framework.}
  \label{Agent Routing}
\end{figure}

\subsubsection{First-Layer Routing}
The first-layer routing is essentially a multi-class classifier with a discrete and finite output space represented as $C = \{c_1, c_2, \ldots, c_k\}$, where $k$ denotes the total number of predefined agent categories. This classifier identifies user query intent $q_i$ and maps it to predefined agent category spaces.


Given a user query $q_i$, the classifier aims to find the most relevant subset of categories:
\begin{equation}
f(q_i) \rightarrow \{c_k \mid c_k \in C, \text{relevance}(q_i, c_k) > \alpha\},
\end{equation}
where $\alpha$ is the relevance threshold, and $\text{relevance}(q_i, c_k)$ represents the relevance score between the query and category $c_i$.

\textbf{Category Design:}
To enhance classification scalability and accuracy, we employ a hybrid classification strategy combining top-down and bottom-up approaches to construct the category system.

\textit{Top-Down Approach}: Based on domain expertise and system architecture planning, we predefine core category sets:
\begin{equation}
C_{\text{predefined}} = \{\text{Image}, \text{Writing}, \text{Travel}, \text{Food}, \text{Shopping}, \text{Finance}, \text{Health}, \text{Education}, \ldots\}.
\end{equation}
These categories possess the domain and semantic characteristics, and encompass major agent application scenarios.


\textit{Bottom-Up Approach} :
Let the agents description set be $A = \{a_1, a_2, \ldots, a_n\}$. For each agent $j$'s functional description  $a_j$, we perform vectorization by a pre-trained SBERT embedding model:
\begin{equation}
\mathbf{e}_j = f_{\text{embed}}(a_j) \in \mathbb{R}^{d},
\end{equation}
where $d$ is the embedding vector dimension.
Then, we employ the K-Means algorithm to cluster agent embedding vectors set $\{\mathbf {e_1, e_2, \cdots, e_n}\}$: 
\begin{equation}
\min \sum_{j=1}^{n} \sum_{\mathbf{e_j} \in S_i} \|\mathbf{e_j} - \boldsymbol{\mu}_i\|^2,
\end{equation}
where $k$ is the number of clusters,      $S_i$ is the $i$-th cluster,     $\boldsymbol{\mu}_i$ is the $i$-th cluster center, and $\|\cdot\|$ is the Euclidean distance. 
Based on clustering results $\{S_1, S_2, \ldots, S_k\}$, we generate data-driven categories $C_{\text{clustered}} $by analyzing common features within each cluster.

The final category system is obtained by merging results from both approaches:
\begin{equation}
C_{\text{final}} = C_{\text{predefined}} \cup C_{\text{clustered}} \setminus C_{\text{redundant}},
\end{equation}
where $C_{\text{redundant}}$ represents redundant categories identified through semantic similarity analysis.
This classification strategy ensures that the category system possesses both theoretical guidance and reflects the distribution characteristics of actual agents, establishing a solid foundation for subsequent precise routing.

\textbf{Category Retrieval via Semantic Similarity.}
In the first-layer routing, the objective is to map an incoming user query $q_i$ to the most relevant subset of categories $\mathcal{C} = \{c_1, c_2, \dots, c_K\}$. We employ a semantic embedding model and cosine similarity to efficiently retrieve the top-$k$ candidate categories.
A pretrained semantic embedding model $f_{\text{embed}}(\cdot)$ is used to map both the query and each category into $d$-dimensional vectors:
\begin{equation}
    \mathbf{q}_i = f_{\text{embed}}(q_i) \in \mathbb{R}^d, \quad 
    \mathbf{c}_j = f_{\text{embed}}(c_j) \in \mathbb{R}^d, \quad j=1,\dots,K.
\end{equation}

Then, for each category $c_j$, we compute the cosine similarity with the query vector:
\begin{equation}
    \text{sim}(\mathbf{q}_i, \mathbf{c}_j) = \frac{\mathbf{q}_i \cdot \mathbf{c}_j}{\|\mathbf{q}_i\| \, \|\mathbf{c}_j\|}, \quad j=1,\dots,K.
\end{equation}
All categories are ranked in descending order according to their similarity scores:
\begin{equation}
    \pi_i = \text{argsort}_{j=1}^K \; \text{sim}(\mathbf{q}_i, \mathbf{c}_j).
\end{equation}
The final output consists of the top-$k$ categories:
\begin{equation}
    \mathcal{C}' = \{c_{\pi_i(1)}, c_{\pi_i(2)}, \dots, c_{\pi_i(k)}\}.
\end{equation}
The algorithm returns $\mathcal{C}'$, the top-$k$ most relevant categories for query $q_i$, which form the candidate search space for the second-layer routing.

\subsubsection{Second-Layer Routing}
The primary aim of the agent second-layer router is to perform a fine-grained selection of one or more agents based on the coarse-grained categories $\mathcal{C'} $ returned by the first-layer router and the original user query $q_i$. Formally, its objective is to map this input to a final, ordered sequence of agents for execution:
\begin{equation}
\mathcal{A}_{\text{selected}} = \mathcal{F}_{\text{router}}\left(q_i, \mathcal{C'}\right).
\end{equation}
We model the agent second-layer router as a \textbf{Context-aware Finite State Machine (CA-FSM)} that adjusts the callability of agents based on context to avoid invoking unavailable agents.
It leverages a hybrid rule-based and semantic reasoning pipeline.
The core decision-making process as a state machine $\mathcal{SM}$, formally represented as a 4-tuple \citep{FSM}:
\begin{equation}
\mathcal{SM} = (S, \Sigma, \delta, \mathcal{A}),
\end{equation}
where: 
\begin{itemize}
    \item $S$: A finite set of states, representing the system's contextual understanding of the query.
    \item $\Sigma$: The input alphabet, comprising user queries $q$ and system events $e$.
    \item $\delta: S \times \Sigma \rightarrow S$: The state transition function.
    \item $\mathcal{A}: S \rightarrow \mathcal{P}(\mathcal{A}_{\text{all}})$: The action function, mapping a state to a subset of the total agent pool ($\mathcal{P}$ denotes the power set), ultimately determining the agents to be invoked.
\end{itemize}

\textbf{State Definitions ($S$):}
The state set $S$ is constructed from atomic and composite states.
    {Atomic States} represent core, singular intents derived from $q$ or $e$:
        \texttt{PATH\_UPLOAD}, \quad \texttt{TRAVEL\_RELATED}, \quad \texttt{FINANCE\_RELATED}, \quad
        \texttt{FOOD\_RELATED}, \quad \texttt{GENERIC\_QUERY}, \quad \texttt{EVENT\_TRIGGERED}.
    {Composite States} represent complex user intents, formed by the conjunction of atomic states: $s_{\text{composite}} = s_1 \cap s_2 \cap \dots \cap s_n$. For example, $\texttt{TRAVEL\_AND\_FOOD}$ indicates a query relevant to both travel and food.

\textbf{State Transitions ($\delta$):} The transition function $\delta(s, \sigma)$ determines the new state based on the current state $s$ and input $\sigma \in \Sigma$. It is implemented via a hybrid mechanism for efficiency and robustness.
\begin{itemize}
    \item \textbf{Rule-Based Pre-Filtering:} A set of lightweight rules $R$ (regex, keyword matching) is applied first to $\sigma$ to assign high-certainty or high-priority states swiftly. Example rule: $\text{IF } \sigma \text{ contains } \text{`/'} \lor \texttt{C:\textbackslash} \lor \text{`upload'} \rightarrow s_{\text{rule}} = \texttt{PATH\_UPLOAD}$.
    This rapidly narrows the candidate agent space.
    \begin{equation}
    s_{\text{rule}} = R(\sigma).
    \end{equation}

    \item \textbf{Embedding-Based Semantic Disambiguation:} For inputs where rules are inconclusive or a composite state is likely, semantic similarity is used. For each atomic state $s_i$, a descriptive text prompt $t_{s_i}$ is defined. Their embeddings $\mathbf{v}_{s_i} = f_{embed}(t_{s_i})$ are precomputed. The input embedding $\mathbf{v}_{\sigma} = f_{embed}(\sigma)$ is calculated. The most probable state is whose embedding vectors are closest to $\mathbf{v}_{\sigma}$, measured by the cosine similarity.
    The resulting semantic state is:
    \begin{equation}
    s_{\text{semantic}} = \underset{s_i \in S}{\arg\max} \, \text{cos}(\mathbf{v}_{\sigma}, \mathbf{v}_{s_i}).
    \end{equation}
    The final state $s_{\text{current}}$ is determined by combining the results of the rule-based and semantic approaches: $s_{\text{current}} = \delta(s, \sigma) = \text{combine}(s_{\text{rule}}, s_{\text{semantic}})$.
\end{itemize}

\textbf{Action Function ($\mathcal{A}$):}
The action function $\mathcal{A}(s_{\text{current}})$ defines the strategy for agent selection given the current state. 
Fetch a relevant subset of agents $\mathcal{A}_{\text{candidates}}$ from the total pool $\mathcal{A}_{\text{all}}$. Firstly, agents are filtered based on naming prefixes derived from $s_{\text{current}}$ and $\mathcal{C'}$. Then, semantic filtering is performed, which is critical for scalability within categories. Let the agent description for agent $a_j$ be $d_{a_j}$. Its embedding $\mathbf{v}_{a_j}$ is precomputed and stored. The query embedding $\mathbf{v}_{q_i} $ is used to perform a similarity search constrained to the agents already filtered by the rule-based step.
        \begin{equation}
        \mathcal{A}_{\text{candidates}} = \underset{a_j \in \{\mathcal{A}_{\text{filtered}}\}}{\arg\max\!\!\!\ _k} \, \text{sim}(\mathbf{v}_{q_i}, \mathbf{v}_{a_j}),
        \end{equation}
        where $\arg\max\!\!\!\ _k$ denotes retrieving the top-$k$ most similar agents.
    
    

After determining the agent's availability by the state machine, the masking strategy presented above is used to improve inference efficiency.

\subsubsection{LLM-Based Final Decision}
The final agent selection from $\mathcal{A}_{\text{candidates}}$ is performed by the LLM, which serves as a powerful ranker and decision-maker. The LLM is provided with a structured prompt $P$ containing the query $q_i$, the current state $s_{\text{current}}$, and the metadata for each agent in $\mathcal{A}_{\text{candidates}}$, which can be formalized as:
\begin{equation}
{a}_{\text{final}} = \text{LLM}\left(P\left(q_i, s_{\text{current}}, \mathcal{A}_{\text{candidates}}\right)\right).
\end{equation}

The LLM is instructed to align the user’s query with the agents’ input parameters while adhering to the contextual constraints defined by $s_{\text{current}}$. Its output is restricted to the final selected agent for invocation.


\subsubsection{KV-cache Based Prefetching Strategy}
Since each query requires two layers of routing inference, redundant computation for duplicate or highly similar queries leads to substantial resource waste. To address this issue, we introduce a high-performance caching strategy designed to reduce latency for frequently occurring queries, lower LLM API costs, and ultimately enhance overall system throughput. 

The proposed prefetching strategy is as follows.
\textbf{Cache Key:} User queries are first standardized by converting to lowercase, removing redundant spaces, and expanding abbreviations. The standardized query is then either directly used as the key or transformed into a semantic embedding.
\textbf{Cache Value:} The cache stores the final list of AI agents to which the query is routed.
\textbf{Process Flow:} Upon receiving a new query, the system first performs a cache lookup. If a cache hit occurs, the stored agent list is returned immediately, bypassing both layers of LLM-based routing. If no match is found, the full routing process is executed, and the resulting output is subsequently written back to the cache for future reuse.

\subsubsection{Adding a new agent}
When a new agent is introduced into the system, the process begins with registration, during which its structured description (\textit{[name, description, input parameters, output parameters]}) is stored in a vector database, together with corresponding few-shot examples to support subsequent routing and inference. 
The system then performs category assignment: embeddings are computed for the names and descriptions of all categories, while the new agent’s description is encoded into an embedding vector. By measuring similarity between the agent embedding and category embeddings, the agent is automatically assigned to the most relevant one or more categories.
If the similarity scores between the new agent and all existing categories fall below a predefined threshold, a new category must be created. 
This can be achieved automatically by a fine-tuned LLM that generates an appropriate category name from the agent’s description.




\subsection{Supervised Fine-Tuning (SFT) based Classification Routing}
\label{APPSFT}
This routing approach formulates the routing problem as a supervised classification task. Given an input prompt $x$, the routing model is trained to predict the most suitable backbone model $m \in \mathcal{M}$, where $\mathcal{M}$ denotes the set of available models. Formally, the routing model learns a mapping function:
\begin{equation}
    f_\theta: x \mapsto m,
\end{equation}
where $f_\theta$ is parameterized by a lightweight neural network, trained using supervised fine-tuning.

Supervised fine-tuning (SFT) plays a crucial role in this approach. Instead of training a router from scratch, we initialize from a pre-trained LLM with strong representation capacity. SFT then adapts the model specifically for the routing task by aligning prompts with their optimal model labels. This not only reduces training cost and improves convergence but also leverages prior knowledge from the pre-training stage to enhance routing performance.

\textbf{{Training Procedure.}}
To construct the training dataset, each prompt $x_i$ is paired with the model $m_i^\ast$ that yields the best response, determined via prior evaluation or human annotation. The dataset can be represented as:
\begin{equation}
    \mathcal{D} = \{(x_i, m_i^\ast)\}_{i=1}^N.
\end{equation}
The routing model is then optimized using the standard cross-entropy loss:
\begin{equation}
    \mathcal{L}(\theta) = - \frac{1}{N}\sum_{i=1}^N \log p_\theta(m_i^\ast \mid x_i),
\end{equation}
where $p_\theta(m \mid x)$ denotes the predicted probability distribution over candidate models. SFT ensures that the router directly learns from explicit supervision, aligning prompts with their most effective models.



\subsection{Contrastive Learning Router Design}
\label{APPCont}
To enhance the routing performance and capture the relative advantages among different candidate models, we design a contrastive learning based router. This approach leverages pairwise supervision signals provided by a strong judge model (we use Gemini 2.5 \citep{2025gemini}) to construct a fine-grained training objective. Specifically, for a given query $x$, we obtain responses $\{r_1, r_2, \dots, r_M\}$ from $M$ candidate models. The judge model evaluates each response along three dimensions, including helpfulness, factuality, and coherence, and produces pairwise preference labels $y_{ij}$, where
\begin{equation}
    y_{ij} = 
    \begin{cases}
        1, & \text{if response $r_i$ is preferred over $r_j$}, \\
        0, & \text{otherwise}.
    \end{cases}
\end{equation}

The router is parameterized as $f_\theta(x, m)$, which outputs a compatibility score between query $x$ and model $m$. For each pair $(i,j)$, we define the probability that model $i$ is preferred over model $j$ as
\begin{equation}
    P(i \succ j \mid x) = \sigma \left( f_\theta(x, i) - f_\theta(x, j) \right),
\end{equation}
where $\sigma(\cdot)$ denotes the sigmoid function. The contrastive loss is then formulated as
\begin{equation}
    \mathcal{L}(\theta) = - \sum_{x} \sum_{i \neq j} \Big[ y_{ij} \log P(i \succ j \mid x) + (1-y_{ij}) \log \big( 1 - P(i \succ j \mid x) \big) \Big].
\end{equation}

This formulation allows the router to learn a relative scoring function that generalizes across models, rather than relying on absolute single-label classification. 
During inference, the router aggregates pairwise predictions to rank all candidate models and selects the most preferred model:
\begin{equation}
    m^{*} = \arg\max_{i} \sum_{j \neq i} \mathbb{I}\big(f_\theta(x,m_i,m_j) > 0\big).
\end{equation}

This design enables the router to capture fine-grained relative strengths and weaknesses among models, leading to strong generalization even when absolute labels are ambiguous. However, its main limitation lies in the high cost of constructing training data, as reliable preference labels depend heavily on the availability of a strong judge model.

\end{document}

%% file: math_commands.tex
\usepackage{amsmath,amsfonts,bm}

\def\eqref#1{equation~\ref{#1}}

\def\1{\bm{1}}

\DeclareMathAlphabet{\mathsfit}{\encodingdefault}{\sfdefault}{m}{sl}
\SetMathAlphabet{\mathsfit}{bold}{\encodingdefault}{\sfdefault}{bx}{n}

